\documentclass[useAMS,usenatbib]{mnras}
\usepackage{epsfig}
\usepackage{graphicx}
\usepackage{epstopdf}
\DeclareGraphicsExtensions{.ps}
\usepackage{times}
\usepackage{natbib}
\usepackage{amssymb}
\usepackage{amsmath}
\usepackage{tabularx}
\usepackage{hyperref}
\usepackage{amsfonts,amsmath,amssymb}
\usepackage{txfonts}
\usepackage{graphicx}
\usepackage{color}
\usepackage{flushend}
\usepackage{enumitem}
\usepackage{tabularx}
\usepackage{color}
\newif\ifAMStwofonts
\AMStwofontstrue
\usepackage{eso-pic}
                                                                                                      
\AddToShipoutPictureBG*{%
  \AtPageUpperLeft{%
    \hspace{0.75\paperwidth}%
    \raisebox{-3.5\baselineskip}{%
      \makebox[0pt][l]{\textnormal{DES 2014-0015}}                                                    
}}}%
                                                                                                      
\AddToShipoutPictureBG*{%
  \AtPageUpperLeft{%
    \hspace{0.75\paperwidth}%
    \raisebox{-4.5\baselineskip}{%
      \makebox[0pt][l]{\textnormal{FERMILAB-PUB-16-132-E}}                                           
}}}%

\newcommand {\aplt} {\ {\raise-.5ex\hbox{$\buildrel<\over\sim$}}\ } 

\newcommand{\be}{\begin{equation}}
\newcommand{\ee}{\end{equation}}
\newcommand{\bes}{\begin{equation*}}
\newcommand{\ees}{\end{equation*}}
\newcommand{\ba}{\begin{eqnarray}}
\newcommand{\ea}{\end{eqnarray}}
\newcommand{\bas}{\begin{eqnarray*}}
\newcommand{\eas}{\end{eqnarray*}}
\newcommand{\brr}{\begin{array}}
\newcommand{\err}{\end{array}}
\newcommand{\bc}{\begin{center}}
\newcommand{\ec}{\end{center}}
\newcommand{\bit}{\begin{itemize}}
\newcommand{\eit}{\end{itemize}}

\newcommand{\msun}{\,{\rm M}_\odot}

\newcommand{\degs}{\mbox{$\textrm{deg}^2$ }}

\newcommand{\lam}{\mbox{$\lambda$}}
\newcommand{\lams}{\mbox{$\lambda \;$}}
\newcommand{\ycyl}{\mbox{$Y_\textrm{500}$}}
\newcommand{\ycyls}{\mbox{$Y_\textrm{500}$}\,}

\newcommand{\yx}{\mbox{$Y_\textrm{X}$}}

\newcommand{\vel}{\,{\rm km\,s^{-1}}}

\newcommand{\mincir}{\raise
  -2.truept\hbox{\rlap{\hbox{$\sim$}}\raise5.truept \hbox{$<$}\ }}
\newcommand{\magcir}{\raise
  -2.truept\hbox{\rlap{\hbox{$\sim$}}\raise5.truept \hbox{$>$}\ }}
\newcommand{\siml}{\raise
  -2.truept\hbox{\rlap{\hbox{$\sim$}}\raise5.truept \hbox{$<$}\ }}
\newcommand{\simg}{\raise
  -2.truept\hbox{\rlap{\hbox{$\sim$}}\raise5.truept \hbox{$>$}\ }}

\newcommand{\rvir}{$R_{500}$}
\newcommand{\mvir}{$M_{500}$}

\title[Optical-SZE Scaling Relations]{Optical-SZE Scaling Relations for DES Optically Selected Clusters within the SPT-SZ Survey}
\author[A. Saro, et al.]{A.~Saro$^{1,2}$, S.~Bocquet$^{1,2}$, J.~Mohr$^{1,2,3}$, E.~Rozo$^{4}$, B.~A.~Benson$^{5,6,7}$, S.~Dodelson$^{5,6,7}$, 
\newauthor E.~S.~Rykoff$^{8,9}$, L.~Bleem$^{6,10}$, T. M. C.~Abbott$^{11}$, F.~B.~Abdalla$^{12,13}$, S.~Allen$^{14,15,16}$, 
\newauthor J.~Annis$^{5}$, A.~Benoit-L{\'e}vy$^{12,17,18}$, D.~Brooks$^{12}$, D.~L.~Burke$^{8,9}$, R.~Capasso$^{1,2}$, 
\newauthor A. Carnero Rosell$^{19,20}$, M.~Carrasco~Kind$^{21,22}$, J.~Carretero$^{23,24}$, I.~Chiu$^{1,2}$, 
\newauthor T.~M.~Crawford$^{6,7}$, C.~E.~Cunha$^{8}$, C.~B.~D'Andrea$^{25,26}$, L.~N.~da Costa$^{19,20}$, S.~Desai$^{1,2}$, 
\newauthor J.~P.~Dietrich$^{1,2}$, A.~E.~Evrard$^{27,28}$, A.~Fausti Neto$^{19}$, B.~Flaugher$^{5}$, P.~Fosalba$^{23}$, 
\newauthor J.~Frieman$^{5,6}$, C.~Gangkofner$^{1,2}$, E.~Gaztanaga$^{23}$, D.~W.~Gerdes$^{28}$, T.~Giannantonio$^{29}$, 
\newauthor S.~Grandis$^{1,2}$, D.~Gruen$^{8,9}$, R.~A.~Gruendl$^{21,22}$, N.~Gupta$^{1,2,3}$, G.~Gutierrez$^{5}$, 
\newauthor W.~L.~Holzapfel$^{30}$, D.~J.~James$^{11}$, K.~Kuehn$^{31}$, N.~Kuropatkin$^{5}$, M.~Lima$^{19,32}$, 
\newauthor J.~L.~Marshall$^{33}$, M.~McDonald$^{34}$, P.~Melchior$^{35}$, F.~Menanteau$^{21,22}$, R.~Miquel$^{24,36}$, 
\newauthor R.~Ogando$^{19,20}$, A.~A.~Plazas$^{37}$, D.~Rapetti$^{1,2}$, C.~L.~Reichardt$^{30,38}$, K.~Reil$^{9}$, 
\newauthor A.~K.~Romer$^{39}$, E.~Sanchez$^{40}$, V.~Scarpine$^{5}$, M.~Schubnell$^{28}$, I.~Sevilla-Noarbe$^{40}$, 
\newauthor R.~C.~Smith$^{11}$, M.~Soares-Santos$^{5}$, B.~Soergel$^{29}$, V.~Strazzullo$^{1,2}$, E.~Suchyta$^{41}$, 
\newauthor M.~E.~C.~Swanson$^{22}$, G.~Tarle$^{28}$, D.~Thomas$^{25}$, V.~Vikram$^{10}$, A.~R.~Walker$^{11}$, A.~Zenteno$^{11}$ }

\begin{document}

\date{\textit{Affilitations are listed at the end of the paper}}

\maketitle                                                 
                                                           
\label{firstpage}      

\begin{abstract} 
We study the Sunyaev-Zel'dovich effect (SZE) signature in South Pole Telescope (SPT) data for an ensemble of 719 optically identified galaxy clusters selected from 124.6~\degs of the Dark Energy Survey (DES) science verification data, 
detecting a clear stacked SZE signal down to richness $\lambda\sim 20$.
The SZE signature is measured using matched-filtered maps of the 2500~\degs SPT-SZ survey at the positions of the DES clusters, and the degeneracy between SZE observable and matched-filter size is broken by adopting as priors SZE and optical mass--observable relations that are either calibrated using SPT selected clusters or through the \citet[][A10]{arnaud10} X-ray analysis.
We measure the SPT signal to noise $\zeta$-\lam\, relation and two integrated Compton-$y$ \ycyl-\lam\, relations for the DES-selected clusters 
and compare these to model expectations that account for the SZE-optical center offset distribution.  
%
For clusters with $\lam > 80$, the two SPT calibrated scaling relations are consistent with the measurements, while for the A10-calibrated relation the measured SZE signal is smaller by a factor of $0.61 \pm 0.12$ compared to the prediction.
For clusters at  $20 < \lam < 80$, the measured SZE signal is smaller by a factor of $\sim$0.20-0.80 (between 2.3 and 10~$\sigma$ significance) compared to the prediction, with the SPT calibrated scaling relations and larger \lam\ clusters showing generally better agreement.
%
%
We quantify the required corrections to achieve consistency, 
showing that there is a richness dependent bias that can be explained by some combination of 
1) contamination of the observables
and 2) biases in the estimated halo masses. 
We discuss also particular physical effects associated to these biases, such as contamination of $\lam$ from line-of-sight projections or of the SZE observables from point sources, larger offsets in the SZE-optical centering or larger intrinsic scatter in the $\lam$-mass relation at lower richnesses.
\end{abstract}


\section {Introduction}
\label{sec:introduction}

Joint analysis of multi-wavelength observations, including optical and Sunyaev-Zel'dovich Effect \citep[hereafter SZE;][]{Sunyaev72}
tracers of the underlying dark matter, will help us in realizing the full cosmological potential of our observational datasets \citep[][]{cunha09,rozo09,song12,rhodes15,bleem15,saro15}.
This joint analysis is particularly important for experiments using galaxy clusters, for which both cluster selection and the determination of mass and redshift generally require multi-wavelength observations \citep[and references therein]{rozo09,mantz10,benson11,bocquet15,mantz15,planck15clusters}.
However, to fully exploit the underlying cosmological information, a careful characterization of possible biases and systematics is crucial, and tests are required to prove consistency among different observable-mass scaling relations associated with different galaxy cluster samples \citep[S15]{rozo14c,rozo14b,rozo14a,evrard14}.

Within this context, several investigations have tested the consistency of optical and SZE properties of galaxy clusters.  
\cite{planck_opt} stacked Planck data at positions of clusters in a catalog selected from Sloan Digital Sky Survey (SDSS) data (the maxBCG catalog, \citealt{koester07}) galaxy clusters and found a deficit of SZE signal compared to what is expected from the associated observed mass-richness and SZE observable-mass scaling relations \citep{johnston07,rozo09,arnaud10}. This result has been confirmed using WMAP data \citep{drapper12}.  In the Planck analysis the discrepancy disappears when using a subset of their optical sample with X-ray emission; however they do not account for the complicated selection function of the X-ray sample they use, and so the resulting agreement could be coincidental.  Similar discrepancies have been noted using data from the Atacama Cosmology Telescope (ACT) in combination with optically selected clusters from the maxBCG catalog \citep{sehgal13} and with a sample of luminous red galaxies \citep{hand11} extracted from the SDSS \citep{kazin10}. On the other hand, a similar SZE stacked analysis based on a sample of locally brightest galaxies have shown more consistent results with respect to the expectation \citep{planck_lbg} down to halo masses of the order of $\sim 4\times 10^{12} \msun$. These apparently discrepant results might be consistent since large theoretical uncertainties are associated with the luminous galaxy halo masses, which in this case have been calibrated through mock catalogs. Results obtained by the \cite{planck_lbg} have been confirmed by \cite{greco15} using a similar sample but with a different analysis method. Consistent results have been also obtained from the stacked-analysis of a sample of radio sources \citep{gralla14}.

From the theoretical perspective, the lower than expected SZE signal for galaxy clusters with a given optical richness 
has been addressed by quantifying and incorporating the systematic uncertainties in the observables as well as the associated covariances \citep{biesiadzinski12,angulo12, rozo14c,evrard14,wu15}.

The aim of this work is to study SZE observables derived from South Pole Telescope (SPT) data at the locations of clusters identified in the Dark Energy Survey (DES) Science Verification Data (DES-SVA1) using the redMaPPer (RM) algorithm \citep{rykoff16} and to test the resulting SZE-optical scaling relations for consistency with the model described in \citet[][hereafter S15]{saro15}.  In S15, a simultaneous calibration of the SZE and optical-richness mass relations is carried out starting from an SPT selected sample of clusters \citep{bleem15} that are matched with the RM cluster catalog extracted from the data of DES-SVA1. We remind the reader that optical richness and SZE scaling relations in S15 were simultaneously calibrated from abundance matching of the SPT-selected sample adopting the same fixed cosmological model used in this work.
 
The plan of the paper is as follows. In section~\ref{sec:data} we describe the RM-selected galaxy cluster catalogs and the SPT data. In section~\ref{sec:theory} we summarize the theoretical model adopted for our analysis.  Section~\ref{sec:sz_observables} describes the method we use to
extract the SZE observables. Our results are presented in section~\ref{sec:results}. Section~\ref{sec:conclusions} contains a discussion of our findings and our conclusions.

For the analysis presented below we adopt fixed cosmology that is flat $\Lambda$CDM with $\Omega_M= 0.3$, $H_0 =70$$\vel$ Mpc$^{-1}$, and $\sigma_8 =0.8$. Cluster masses (\mvir) are defined within \rvir, the radius within which the average density is 500 times the critical density of the Universe at the cluster redshift.

\begin{figure*}
\centering { \hbox{
    \includegraphics[width=60mm,height=45mm]{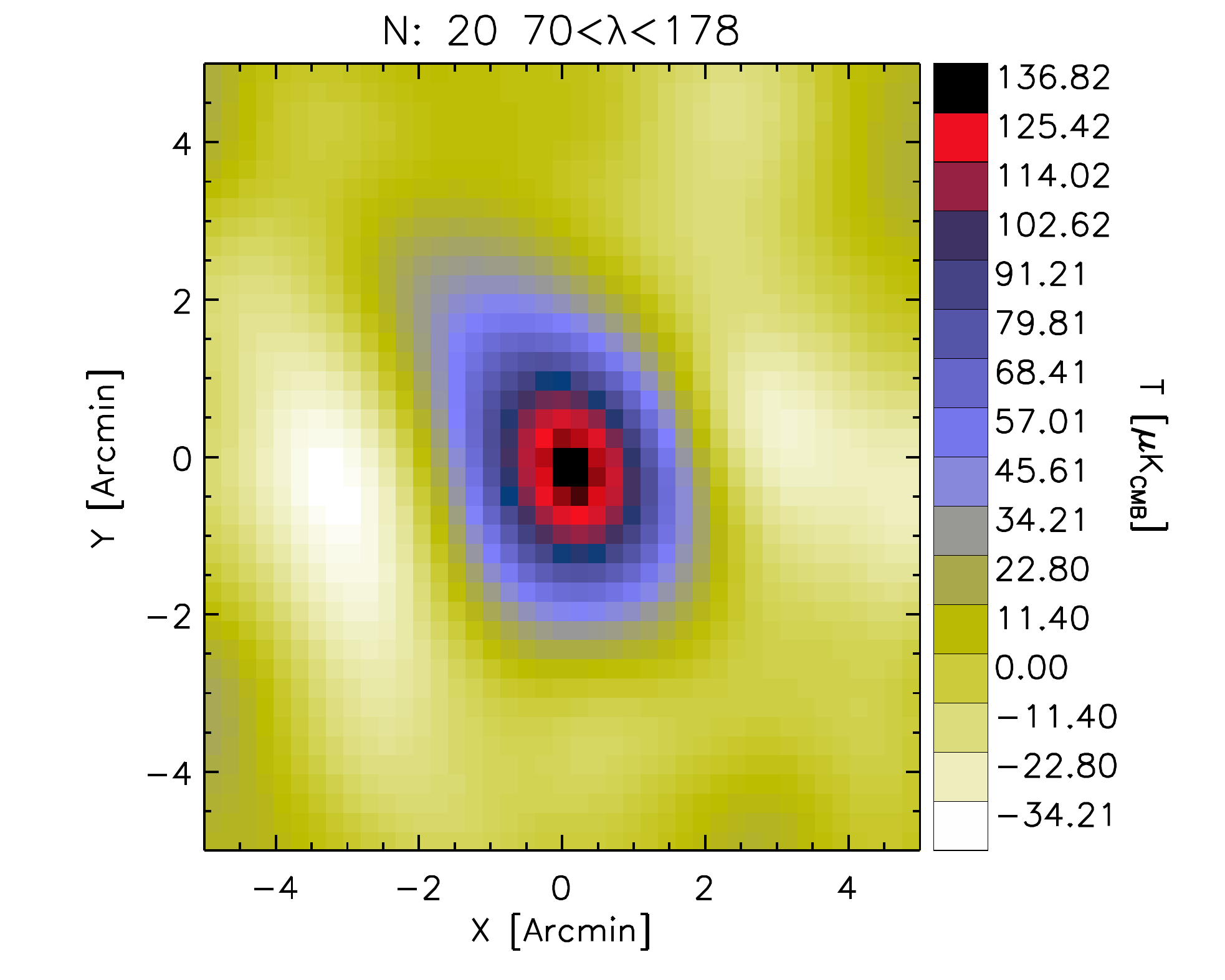}
    \includegraphics[width=60mm,height=45mm]{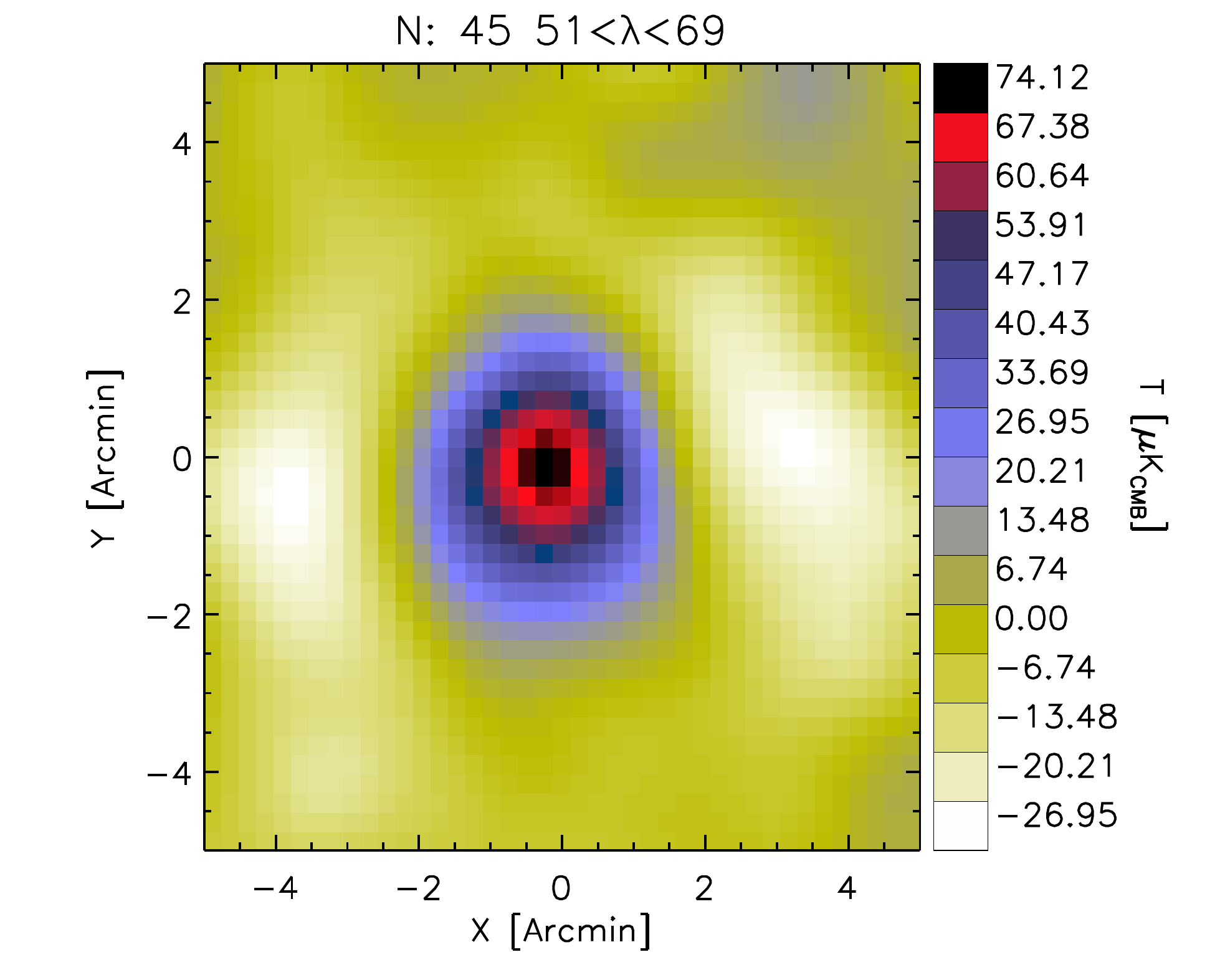}
    \includegraphics[width=60mm,height=45mm]{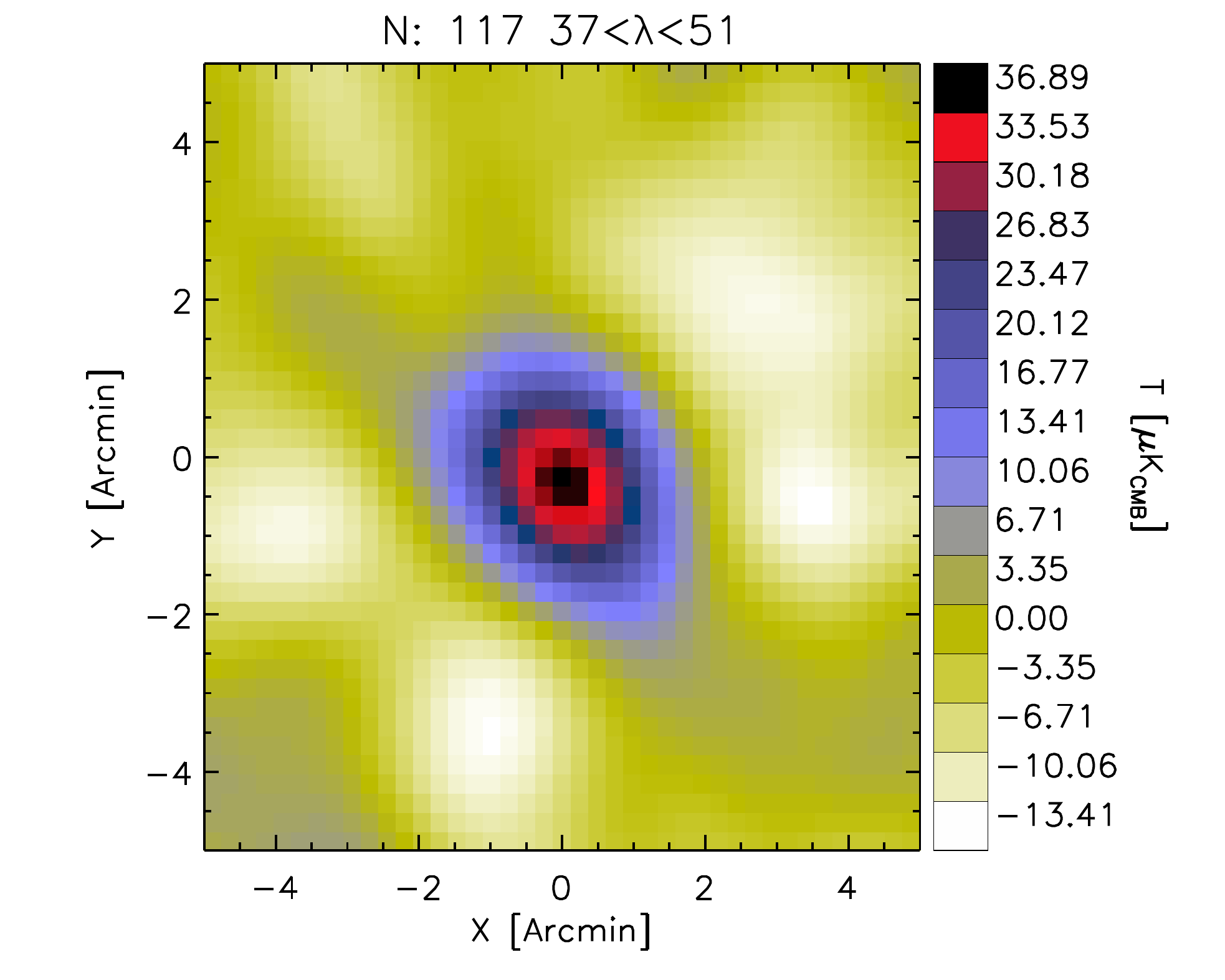} } }
\centering { \hbox{
    \includegraphics[width=60mm,height=45mm]{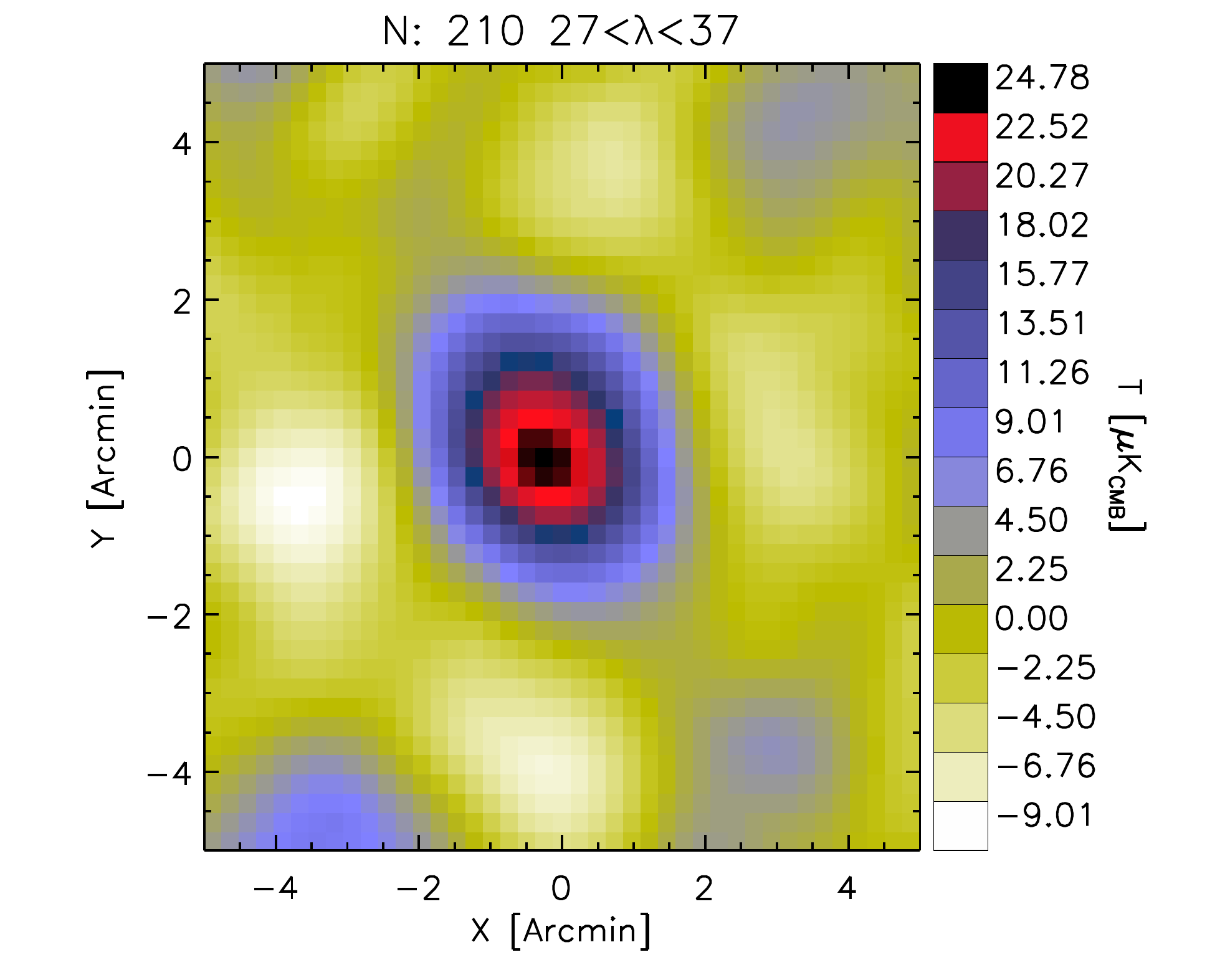}
    \includegraphics[width=60mm,height=45mm]{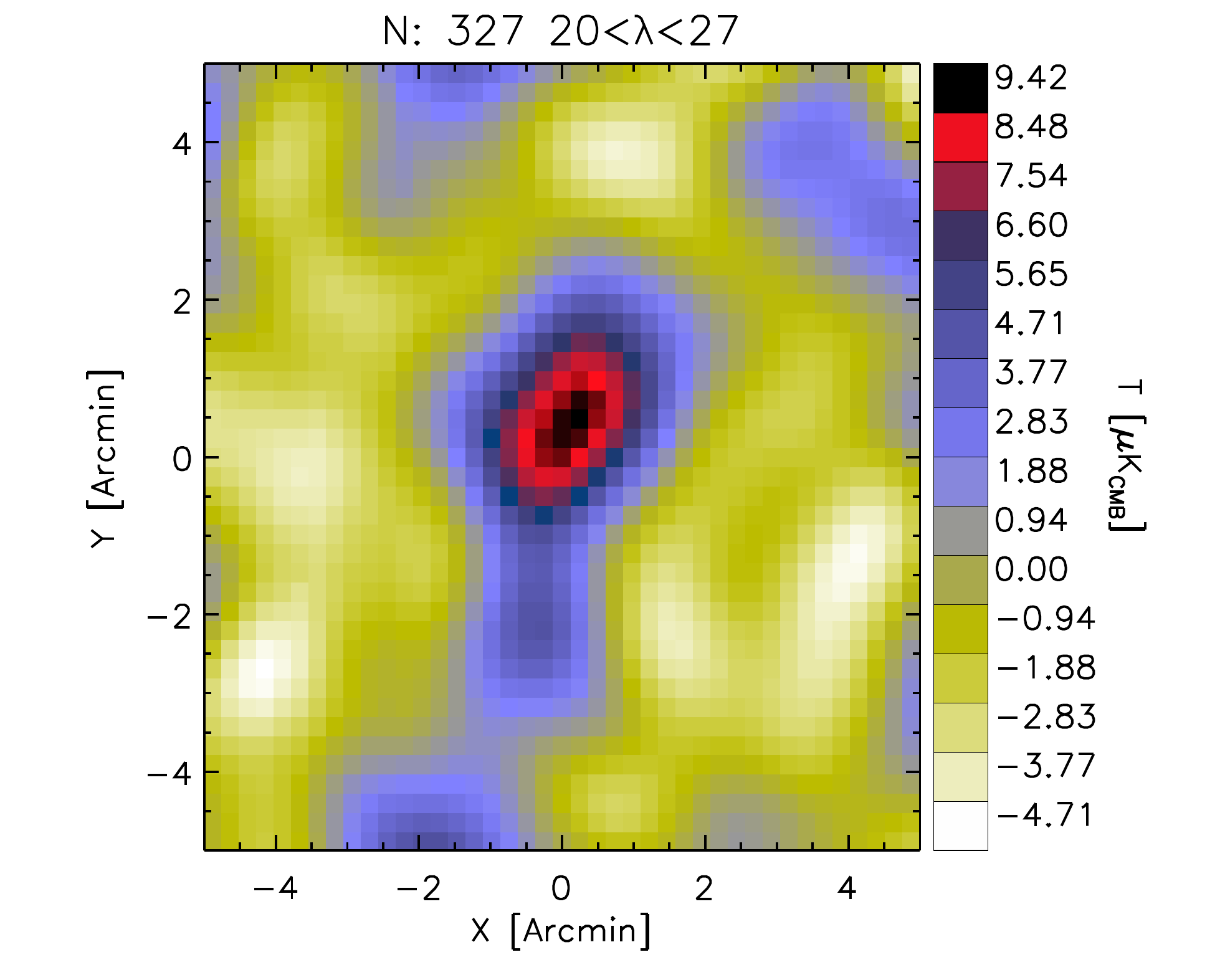}
    \includegraphics[width=60mm,height=45mm]{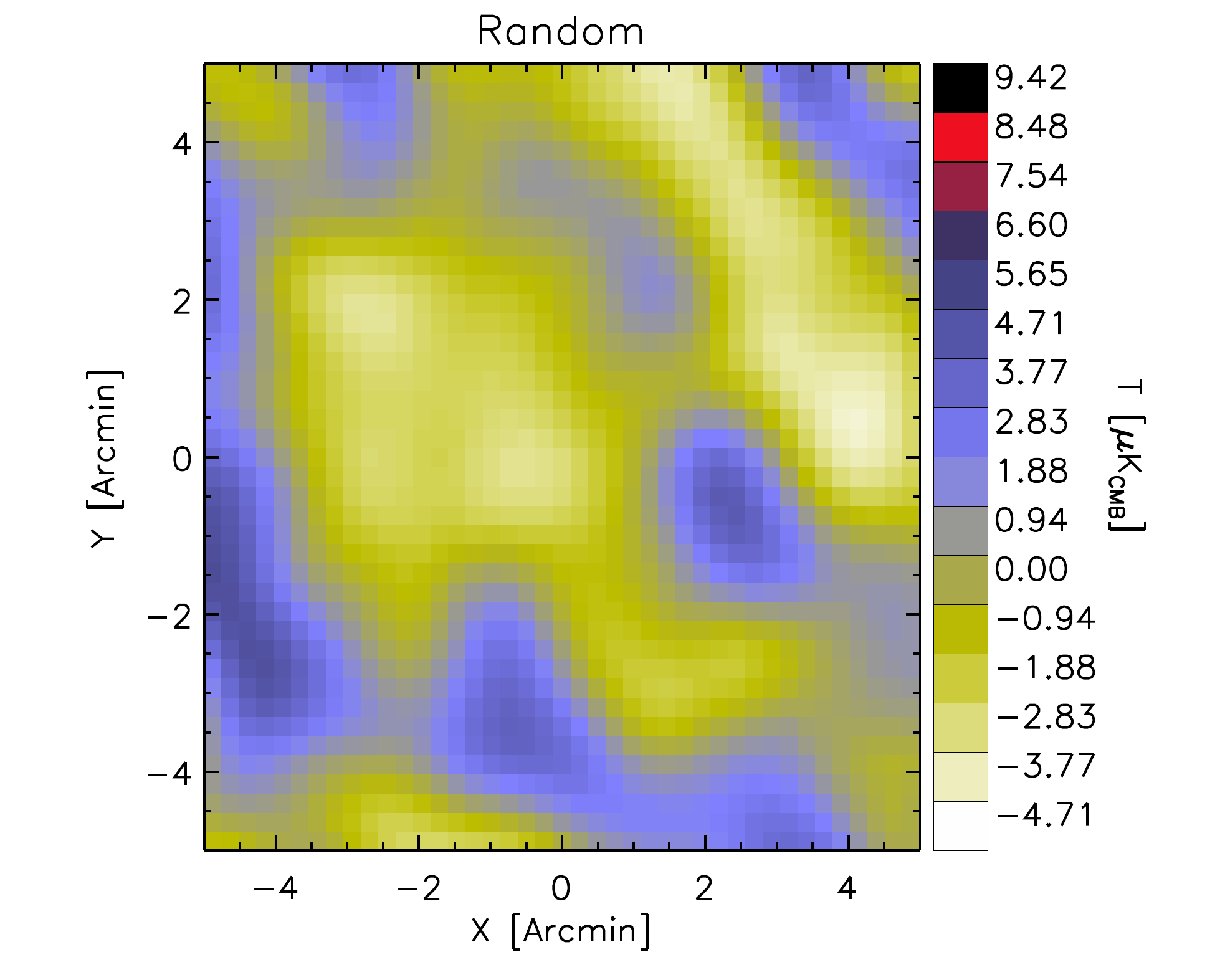} } }
\caption{ From top-left, the first 5 panels show the temperature decrement of the stacked SPT filtered maps centered at the position of RM-selected clusters in bins of decreasing \lam. For each RM cluster the SPT maps have been filtered with an SZE profile using a mass derived from the \lam-mass relation in S15. The number of clusters and \lam \, range for each bin are listed above each panel. The bottom-right-most panel shows the same SPT filtered map shown on the bottom-central panel but with random sky positions. Note the colour scale in the maps changes from map to map, except for the two right-most plots in the bottom row.}
\label{fig:stack_maps}
\end{figure*}

\section{Data}
\label{sec:data}

\subsection{DES Optical Cluster Catalogs}
\label{sec:des}

For this work we use optically selected clusters from the DES-SVA1 region that overlaps with the SPT-SZ 2500d survey.  In section~\ref{subsec:des} we describe the acquisition and preparation of the DES-SVA1 data, and in section~\ref{subsec:redmapper_clusters} we describe the RM cluster catalog. 

\subsubsection{DES-SVA1 Data}
\label{subsec:des}
The DES-SVA1 data include imaging in grizY of $\sim300$~\degs over multiple disconnected fields \citep[e.g.][]{rykoff16},
most of which overlap with the SPT-SZ survey. The 
DES-SVA1 data were acquired with the Dark Energy Camera \citep{flaugher15} over 78 nights, starting in November 2012 and ending in February 2013 with depth comparable to the nominal depth of the full DES survey.

Data have been processed through the DES Data Management system (DESDM) that is an advanced version of development versions described in several earlier publications \citep{ngeow06,mohr08,mohr12,desai12}.  The coadd catalog produced by DESDM of the SV dataset was analyzed and tested, resulting in the generation of the SVA1 \textit{Gold} catalog \citep{rykoff16}.
The \textit{Gold} catalog covers $\sim250$~\degs and is optimized for extragalactic science. In particular, it avoids the Large Magellanic Cloud and its high stellar densities and is masked south of declination $\delta=-61^\circ$. Furthermore, the footprint is restricted to the regions where data have coverage in all of griz \citep{rozo15}.

\subsubsection{redMaPPer Cluster Catalog}
\label{subsec:redmapper_clusters}

The red-sequence Matched-Filter Probabilistic Percolation (redMaPPer) algorithm is a cluster-finding algorithm based on the improved richness observable of \citet{rykoff12}. RM has been applied to photometric data from the SDSS Stripe 82 coadd data \citep{annis14}, to the the Eighth Data Release (DR8) of SDSS \citep[][]{aihara11} and to DES-SVA1 and DES Year 1 data \citep[][S15]{rykoff16,soergel16}, and has been shown to provide excellent photometric redshifts, richness estimates that tightly correlate with external mass proxies \citep[S15,][]{rykoff16}, and very good completeness and purity \citep{rozo14b,rozo14d, rozo14a}. 

We employ an updated version of the RM algorithm (v6.3.3), with improvements presented in \citet{rozo14b}, \citet{rozo15}, and \citet[]{rykoff16}. The main features of the algorithm can be summarized as follows:
(1) Colours of red-sequence galaxies are calibrated using galaxy clusters with spectroscopic redshifts. (2) Red-sequence galaxy calibrated colours are then used to estimate the membership probability for every galaxy in the vicinity of a galaxy cluster candidate. (3) The richness $\lambda$ is then defined as the sum of the membership probabilities ($p_{\mathrm{RM}}$) over all galaxies:
\begin{equation}
\lambda = \sum p_\textrm{RM}. 
\end{equation}
(4) The RM centering algorithm is probabilistic and the centering likelihood incorporates the following features:  $i$) The photometric redshift of the central galaxy must be consistent with the cluster redshift, $ii)$ the central galaxy luminosity must be consistent with the expected luminosity of the central galaxy of a cluster of the observed richness, and $iii)$ the galaxy density on a 300 kpc scale must be consistent with the galaxy density of central galaxies. The centering probability $P_{cen}$ further accounts for the fact that every cluster has one and only one central galaxy. 
A prior is placed that the correct central galaxy of a cluster is one of the 5 most likely central galaxies in the cluster field. This prior is uninformative while still limiting the number of candidate centrals for which $P_{cen}$ must be estimated.

The DES-SVA1 RM catalog \citep{rykoff16} was produced by running on a smaller footprint than that for the full SVA1 \textit{Gold} sample. In particular, we restrict the catalog to the regions where the $z$-band $10\sigma$ galaxy limiting magnitude is $z>22$. In total, we use $148\,\mathrm{deg}^2$ of DES-SVA1 imaging, with $129.1\,\mathrm{deg}^2$ overlapping the SPT-SZ footprint.  In this area, the largest fraction ($124.6\,\mathrm{deg}^2$) is included in the so called DES-SVA1 SPT-E field. The final catalog used in this work consists of 719 clusters with $\lambda > 20$ and redshifts in the range $0.2<z<0.9$.

\subsection{SPT-SZ 2500d survey}
The thermal SZE signals analyzed in this work are extracted at 95 and 150 GHz from the 2500d SPT-SZ survey. (For a description of the survey, see, e.g., Bleem et al. 2015). We use point-source masked maps, which mask an area around each point source detected at more than $5\sigma$ in the 150 GHz data, which gives a total useable area of 2365 deg$^2$. Typical instrumental noise is approximately 40~(18) $\mu {\rm K_{CMB}}$-arcmin and the beam FWHM is 1.6~(1.2)~arcmin for the 95~(150)~GHz maps. The SPT-SZ survey is divided into 19 different sub-fields, whose relative noise levels can be characterized by the relative field scaling factor $A_s $. 
We refer the reader to \cite{bleem15} and to \cite{schaffer11} for further details on the SPT-SZ survey and map-making.
We use a multi-frequency matched filter to extract the thermal SZE signal from clusters and enhance the signal-to-noise. This approach is designed to optimally measure the cluster signal
given knowledge of the cluster profile and the noise in the maps  \citep{haehnelt96,melin06}. Therefore, additional information to describe the shape of the cluster profile is required, while we leave the amplitude of model free. In this study, the cluster pressure profiles are assumed to be well fit by the X-ray derived A10 universal pressure profile model \citep{plagge10}:
\be
P(x) \propto \left\{ (1.177 x)^{0.31} [1+(1.18x)^{1.05}]^{4.93} \right\}^{-1},
\label{eq:mf}
\ee
where $x=r/$\rvir.

We thus describe the adopted profiles in terms of the projected radius $\theta_{500}$ and adopt 30 different profiles linearly spaced in $\theta_{500}$ from 0.5 to 7.75~arcmin. 
The noise contributions are both astrophysical (e.g., the CMB, point sources) and instrumental (e.g., atmospheric, detector) based.  We adopt the following Fourier domain spatial filter:
\begin{equation}                                                                
\label{eqn:filter}                                                              
\psi(k_x,k_y) = \frac{B(k_x,k_y) S(|\vec{k}|)}{B(k_x,k_y)^2
  N_{astro}(|\vec{k}|\ ) + N_{inst}(k_x,k_y)}
\end{equation}                                                                  
where $\psi$ is the matched filter, $B$ is the SPT beam and $S$ is the
assumed source template. The noise contributions $N_{astro}$ and
$N_{inst}$ respectively encapsulate the astrophysical and
the instrumental noise.

\section{Theoretical Framework}
\label{sec:theory}

Neglecting small relativistic corrections, the thermal SZE in the direction of a cluster at a frequency $\nu$ can be approximated by (\citealt{sunyaev80b}):  $\Delta
T(\nu) \simeq T_\textrm{CMB} G(\nu)y_{\rm{c}}.$
Here $T_\textrm{CMB} $ is the cosmic microwave background (CMB) temperature and $G(\nu)$ describes the thermal SZE frequency dependence. The Comptonization parameter $y_c$ is related to the integrated pressure along the line of sight:
\be
y_{\rm{c}}=(k_{\rm{B}}\sigma_{\rm{T}}/m_{\rm{e}}c^2)\int n_{\rm{e}}T_{\rm{e}} dl,
\label{eq:yc}
\ee
where $n_{\rm{e}}$ and $T_{\rm{e}}$ are the electron density and temperature, respectively.

In this work we study two different SZE observables: the SPT detection significance $\zeta$ and the Comptonization parameter $y_c$ integrated over the solid angle $\theta_{500}$ defined as $Y_{500}=\int_{0}^{\theta_{500}} y_c \textrm{d}\Omega$, where $\theta_{500}$ is the projected angle associated to \rvir. For the latter we analyse two different models that use different \ycyl-mass relations. We will relate both SZE observables $\zeta$ and \ycyl\ to the central decrement $y_0$, as discussed in details in Sec. \ref{sec:sz_observables}.

\subsection{$\zeta$-mass Relation from SPT}

Following previous SPT analyses (\citealt[][]{benson11, bocquet15, bleem15}; S15), we describe the $\zeta$-mass relation as a log-normal distribution with scatter $D_{\textrm{SZE}}$ and a mean:
\begin{eqnarray}
 \langle \textrm{ln} [(1-f)\zeta]| M_{500}, z \rangle&=& \textrm{ln}
A_\textrm{SZE} + {B_{\textrm{SZE}}}\, \textrm{ln} \left ( \frac{(1-b) M_{500}}{3
  \times10^{14} h^{-1} \msun} \right ) \nonumber \\
  & & +\, {C_{\textrm{SZE}}}\, \textrm{ln}
\left ( \frac{E(z)}{E(z=0.6)} \right ),
\label{eq:zeta}
\end{eqnarray}
where $E(z)\equiv H(z)/H_0$, with best fitting parameters $A_\textrm{SZE} = 4.02\pm 0.16$, $B_\textrm{SZE}= 1.71\pm 0.09$, $C_\textrm{SZE} = 0.49 \pm 0.16$ and $D_\textrm{SZE}=0.20\pm 0.07$ as given in S15. 
Here, we introduce two parametrizations to account for possible biases in mass ($1 - b$) and in the SZE observable ($1 - f$). While we note that the two parameters are mathematically equivalent (besides a trivial transformation with the slope of the scaling relation $B_\textrm{SZE}$), in the following we will keep a formal distinction between $b$ and $f$ as the physical causes for each term are quite different. We also note that only a constant fractional contamination in the optical observable that is independent of richness and redshift can be modelled by a constant fractional bias in mass ($b$) or in the SZE observable ($f$).

\subsection{\ycyl-mass Relation from A10}

For the first model of the spherical $Y$-mass scaling relation, we follow \citet{arnaud10} (hereafter A10) and adopt a log-normal distribution with intrinsic scatter $\sigma_Y = 17\%$ \citep{planck20} and mean:
\ba
\label{eq:arnaud10sc}
\langle \textrm{ln} [(1-f)Y^\textrm{sph}_{500 } /  \textrm{Mpc}^2 ]|  M_{500}, z \rangle &=&  \textrm{ln}(10^{-4.739} h_{70}^{-5/2} E(z)^{2/3}) \,+ \nonumber\\
&& 1.79 \textrm{ln} \left (\frac{(1-b) M_{500}}{3\times 10^{14}h_{70}^{-1} \msun} \right), \ea 
where the cylindrical and spherical quantities are related as $\ycyl=1.203 Y^\textrm{sph}_{500}$ and where the bias parameters $b$ and $f$ are equal to zero as in A10.

\subsection{\ycyl-mass Relation from SPT}

As an alternative, for the second model of the \ycyl-mass scaling relation we adopt the form derived using the \yx-mass calibration of the SPT cluster sample.
Following \cite{andersson11} we assume $Y^\textrm{sph}_{500 }=0.92 \yx$ and describe $\langle \yx | M_{500},z\rangle $ as:
\ba
\label{eq:yx}
&& \langle \textrm{ln} [(1-f)Y_\textrm{X} / 3\times 10^{14} \msun \textrm{keV}] |M_{500} ,z \rangle =  \left[ \textrm{ln} \left( \frac{(1-b) M_{500}}{3\times 10^{14}M_\odot} \right) - \right. \nonumber \\ &&
\left. \textrm{ln} \left( A_\textrm{X}h^{1/2}\right) - \frac{5B_\textrm{X}-3}{2}\textrm{ln}\left( \frac{h}{0.72} \right) -C_\textrm{X}\textrm{ln}E(z) \right]{B_\textrm{X}}^{-1}.
\ea
For consistency, we fit the \yx-mass relation using \yx\ measurements from 82 SPT-selected clusters \citep{mcdonald13} as described in detail elsewhere \citep{dehaan16} 
under the same assumption adopted in S15. Note that the adopted functional form describing the \yx-mass relation in \cite{dehaan16} is inverted with respect to the other observable-mass relations used in this work.
We obtain $A_\textrm{X} = 6.7 \pm 0.37$, $B_\textrm{X} = 0.43 \pm 0.03$ and $C_\textrm{X} = -0.12 \pm 0.14$ for bias parameters $b$ and $f$ equal to zero. Note that, as for the $\zeta$ and \lam\ mass relations, the \yx\ parameters in this work have been derived from abundance matching of the SPT-selected cluster sample assuming the same fixed reference cosmology. They therefore differ from the ones presented in \citet{dehaan16}, which are simultaneusly calibrated within a cosmological analysis. Our approach allows us to make a direct consistency check among the different scaling relations.
We also note that, as a result, the associated slope of the \ycyl-mass relation in this case ($1/B_\textrm{X} = 2.32$) is much steeper ($\sim2.5\sigma$) than the corresponding slope in the A10 relation (1.79). This is an outcome of the SPT cluster abundance calibration \citep[for discussion, see][]{dehaan16}.

\subsection{$\lambda$-mass Relation Calibrated with SPT}
On the optical side we assume that the RM-selected sample is also well described by 
the \lam-mass relation calibrated in S15 for the SPT selected sample
. Namely we assume a \lam-mass relation of the form:
\begin{eqnarray}
\langle \textrm{ln} \lambda| M_{500}, z \rangle & = &  \textrm{ln} A_\lambda + B_\lambda\, \textrm{ln} \left ( \frac{M_{500}}{3\times10^{14} h^{-1}\msun} \right ) \nonumber \\
& & + C_\lambda\, \textrm{ln} \left ( \frac{E(z)}{E(z=0.6)} \right ).
\label{eq:lambda_scaling}
\end{eqnarray}
An additional parameter $D_\lambda$ describes the intrinsic scatter in $\lambda$ at fixed mass, which is assumed to be log-normal and uncorrelated with the SZE scatter, with variance given by:
\be
\textrm{Var}(\textrm{ln} \lambda|M_{500}) = \textrm{exp} ( - \langle \textrm{ln} \lambda|M_{500} \rangle ) + D_\lambda^2.
\label{eq:lambda_scatter}
\ee 
Best fit parameters (and associated uncertainties) of the \lam-mass relation are taken from S15 and are  $A_\lambda = 66.1_{-5.9}^{+6.3}$, $B_\lambda= 1.14_{-0.18}^{+0.21}$, $C_\lambda = 0.73_{-0.75}^{+0.77}$ and $D_\lambda = 0.15_{-0.07}^{+0.10}$. We remind the reader that the parameters of the \lam-mass and of the $\zeta$-mass relations have been simultaneously calibrated in S15 from an SPT-selected cluster sample. Thus, a consistency of these scaling relations 
(which implies $b$ and $f$ consistent with zero)
 is expected for the high-richness end of the cluster population, corresponding approximately to the region $\lam>80$ (S15).

\subsection{Mass-prior and Miscentering}
Following \cite{liu15} and \cite{bocquet15}, for every RM cluster selected with richness $\lambda \pm \Delta \lambda$, at a redshift $z$, we compute the mass prior\footnote{In the following we will refer to \mvir\ as $M$.}: 
\be
P(M|\lambda,z) \propto P(\lambda|M,z) P(M,z),
\label{eq:p_m_l}
\ee
 where $ P(\lambda|M,z)$ is obtained by convolving the scaling relation described by equation~(\ref{eq:lambda_scaling}) and (\ref{eq:lambda_scatter}) with the associated Gaussian measurement uncertainty $\Delta \lambda$, and $P(M,z)$ represents the mass prior and is proportional to the halo mass function \citep{tinker08}. For every cluster in the sample, $P(M|\lambda,z)$ is then marginalized over the \lam-mass scaling relation parameter uncertainties. We then use the resulting mass prior $P(M|\lambda,z)$ in equations~(\ref{eq:zeta}), (\ref{eq:arnaud10sc}) and (\ref{eq:yx}) to compute the expected SZE observable of each RM-selected cluster.

When exploring the impact of miscentering on our observations, we assume the RM-selected sample is characterized by the optical-SZE central offset distribution calibrated in S15 from the SPT selected sample.  The impact of the corresponding corrections is discussed in section~\ref{sec:misc}.

For the purpose of this analysis, results presented in this work have been obtained by fixing the SZE observable-mass scaling relation parameters to their best fit values, while we marginalize over the uncertainties of the \lam-mass relation parameters and over the parameters of the adopted optical-SZE central offset distribution. This approximation is justified because the uncertainties of the parameters describing the \lam-mass relation are 
dominated by uncorrelated Poisson errors and significantly larger
than the ones describing the SZE observable-mass scaling relations.

\section{SZE observables}
\label{sec:sz_extraction}
Using the matched-filter (equation~\ref{eqn:filter}), we extract the central decrement temperature $T_0$ = $T_\textrm{CMB} y_0$, where $y_0$ is the filter amplitude and represents the central Comptonization parameter for every RM-selected cluster.  In addition, we extract the associated (Gaussian) measurement uncertainty $\Delta_{T_0} = T_\textrm{CMB}  \Delta_{y_0}$.

As an example, $5\times5$~arcmin stacked SPT filtered maps showing the central average $T_0$ decrement in five bins of decreasing richness \lams are shown in Fig.~\ref{fig:stack_maps}.  The number of stacked clusters and \lams range for each bin are listed above each panel. The bottom-right panel shows the same SPT filtered map shown on the bottom-central panel but stacked at random sky positions.   Each of these maps has been obtained by stacking the SPT maps at the location of each RM cluster, matched-filtered assuming a cluster profile with mass $\langle M_{500}|\lambda,z,\rangle$ according to equation~(\ref{eq:p_m_l}) and the parameters of the \lam-mass scaling relation derived in S15.
\begin{figure*}
\centering { 
\hbox{
    \includegraphics[width=85mm]{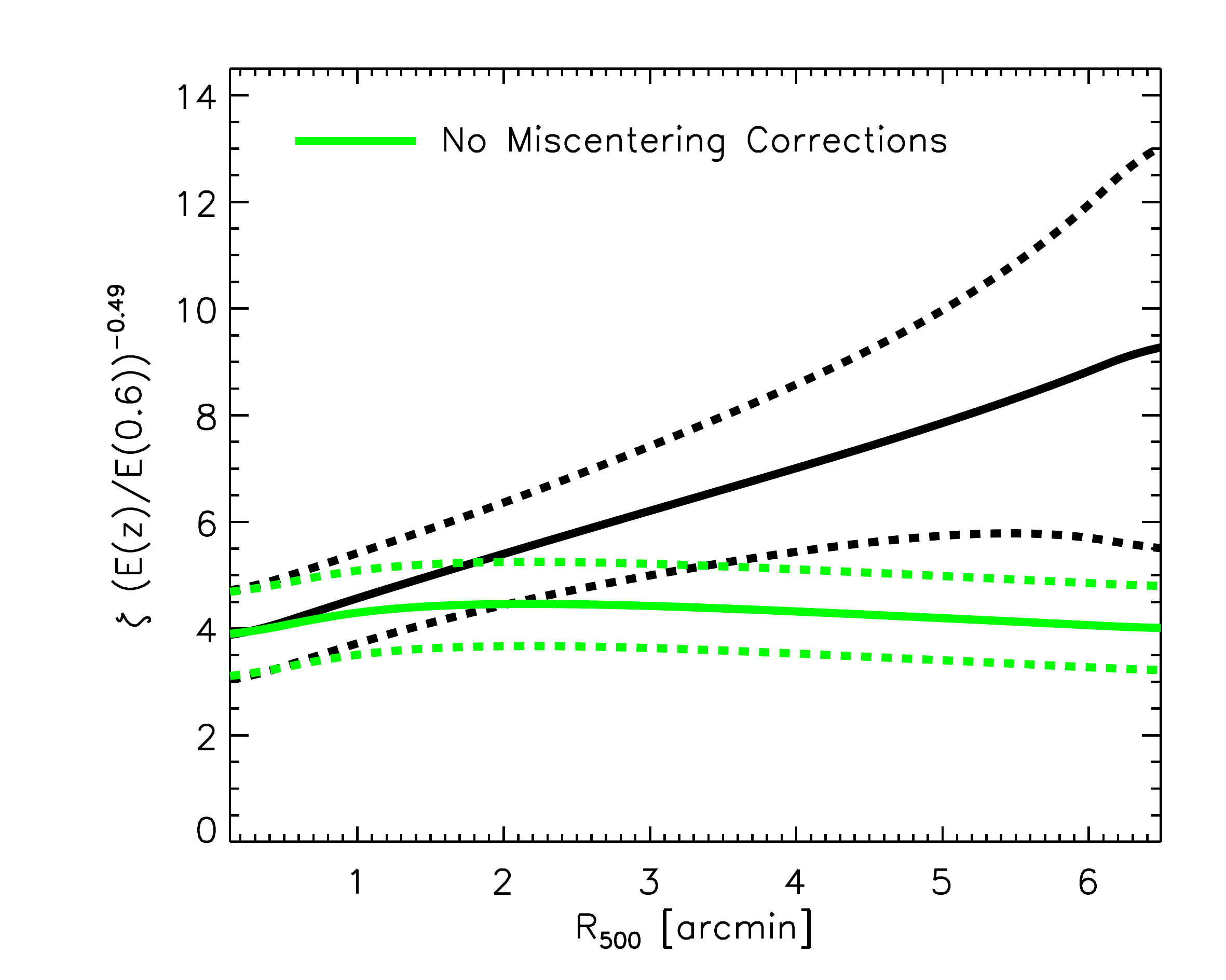}
    \includegraphics[width=85mm]{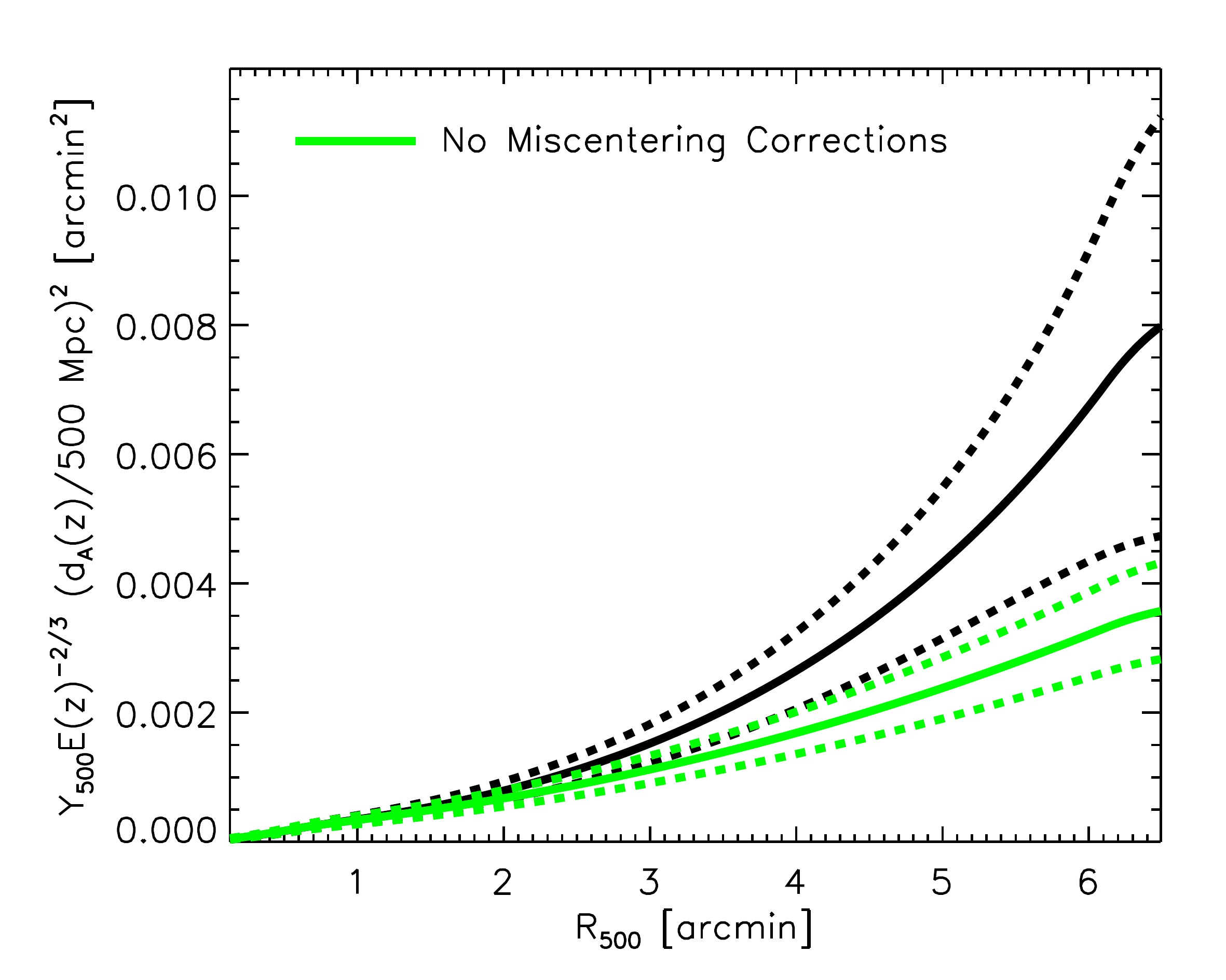}}
}
\caption{The redshift-evolution-corrected distribution 
of the SZE observables $\zeta$ (left panel) and \ycyls (right panel) as a function of cluster extent \rvir\ measured in the SPT filtered map for a RM-selected cluster with $\lambda=77.8 \pm 5.3$ at $z=0.4$ (also SPT-detected as SPT-CL J0447-5055 with raw signal-to-noise of $5.97$). Green (black) solid and dotted lines show the measured $\zeta$ and \ycyls distributions and $1\sigma$ measurement uncertainties without (with) optical-SZE central offset corrections (see section~\ref{sec:sz_observables}).  Note that corrections from the adopted optical-SZE central offset model dominate the degeneracy between $\zeta$ and cluster extent.}

\label{fig:multi_1}
\end{figure*}

\begin{figure*}
\centering { 
\hbox{
    \includegraphics[trim=-30 0 30 0mm,height=85mm]{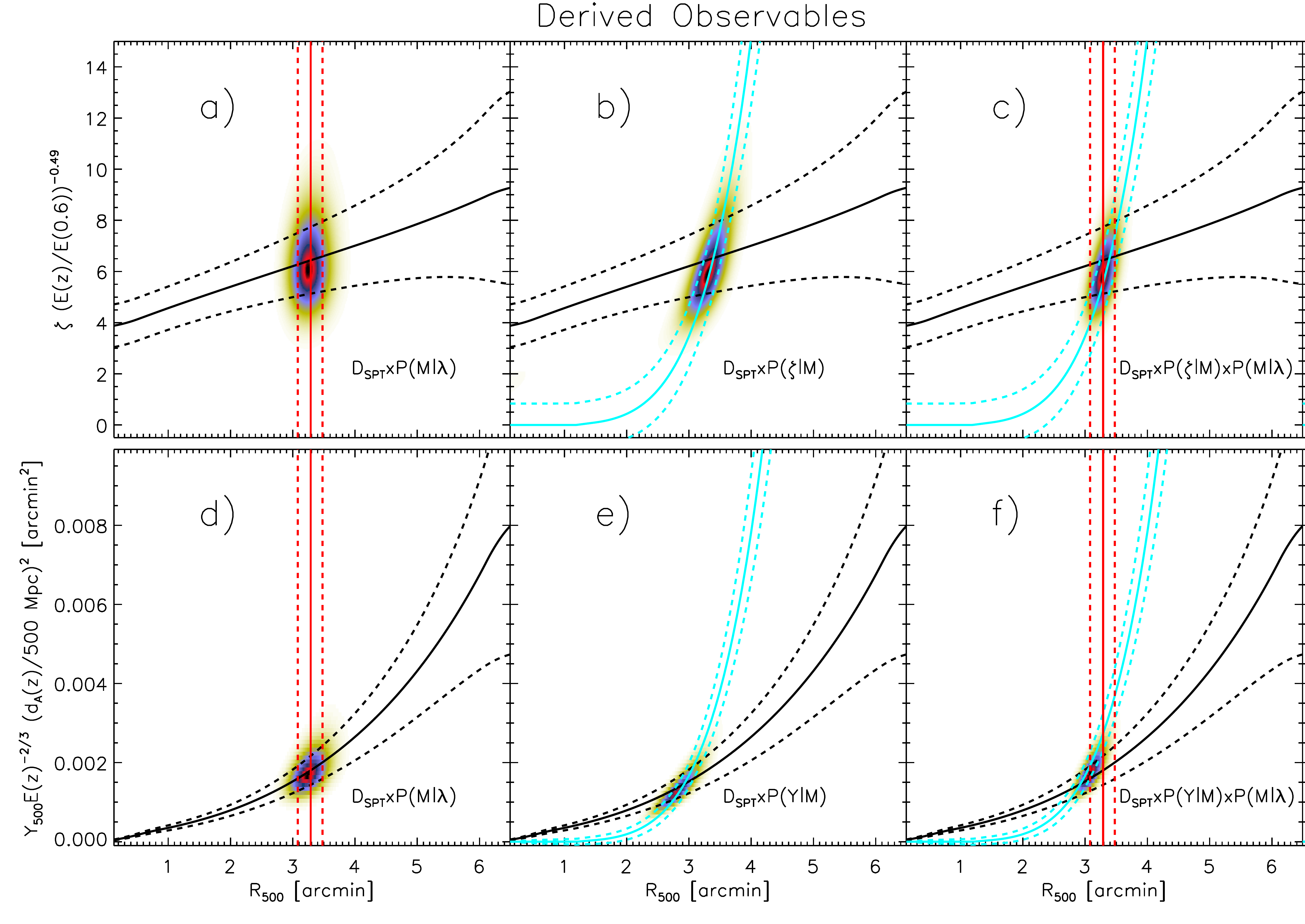}
    \includegraphics[trim=-90 0 90 0mm,height=85mm]{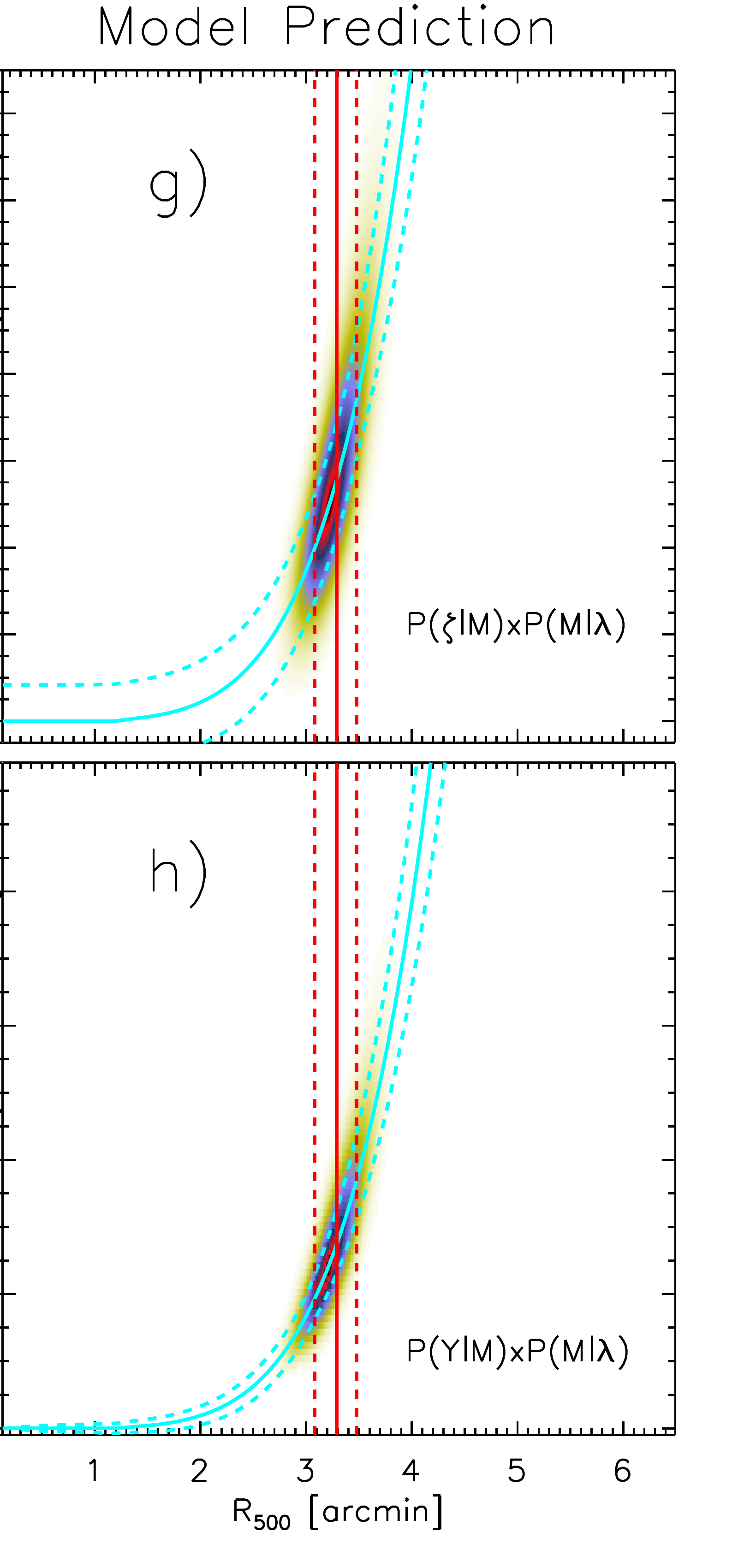}
}
}
\caption{\textit{First three columns:} 
The probability distributions of the redshift-evolution-corrected SZE observables $\zeta$ (upper panel) and \ycyls (lower panel) measured from the SPT filtered maps as a function of cluster extent for the same cluster shown in Fig.~\ref{fig:multi_1}. Distributions are shown for the case of the SPT data (black lines) including 1) the richness-mass prior (red lines, first column); 2) the $\zeta$ and A10 \ycyl\ mass priors (cyan lines, second column) and 3) both priors (third columns). \textit{Last column:} The predicted distribution of the SZE observables for the given richness-mass and SZE observable-mass model.
}
\label{fig:multi_2}
\end{figure*}

\label{sec:sz_observables}
In this work we study two different SZE observables: the detection significance $\zeta$ and \ycyl, the integrated Compton-$y$ within \rvir.  For every RM-selected system, we start from the measured $y_0 \pm \Delta {y_0}$ at the cluster location extracted from the SPT filtered-maps for each assumed cluster profile $\theta_{500} \in [0.5 - 7.75 \, \textrm{arcmin}]$.
We then compute the expected bias $\beta$ associated with each optimal filter due to the optical-SZE central offset distribution by marginalizing over the parameters derived in S15. 
We therefore calculate the estimated average miscentering-corrected central decrement $y_{0,\textrm{corr}} = y_0 / \beta$.  The SZE observables $\zeta$ and \ycyl\ are then related to the miscentering-corrected central decrement $y_{0,\textrm{corr}}$ as follows.
The SPT unbiased significance is defined as $\zeta = y_{0,\textrm{corr}}/(\Delta {y_0} A_s) $ with Gaussian measurement uncertainty $\Delta {\zeta} = A_s^{-1}$, where $A_s$ is the relative field scaling factor \citep{bleem15}.
The integrated \ycyl\ is defined as $\ycyl=\mu y_{0,\textrm{corr}}$, where we numerically integrate the assumed A10 universal pressure profile along the line-of-sight to obtain the scaling the factor $\mu$ \citep{liu15}.

\subsection{\mbox{Degeneracy between Observable and Cluster Extent: $D_\textrm{SPT}$}}
\label{sec:dspt}
As a representative example, we show in the left panel of Fig.~\ref{fig:multi_1} the derived SPT observable $\zeta$ estimated from the SPT filtered maps as a function of assumed cluster extent for a RM-selected cluster identified at redshift $z=0.40 \pm 0.01$ with $\lambda=77.8 \pm 5.3$. This cluster is also identified in the 2500d SPT-SZ survey as SPT-CL J0447-5055, and presented in \cite{bleem15} with with raw signal-to-noise of $5.97$ 
and angular scale $\theta_{500} =  3.05$~arcmin. Green lines represent the mean (solid line) and $1\sigma$ measurement uncertainties (dotted lines) associated with $\zeta$ without the SZE-optical central offset correction, while the mean and $1\sigma$ measurement uncertainty in the case including the SZE-optical central offset correction are shown using black lines. The same distribution is reported in black in the upper panels of Fig.~\ref{fig:multi_2}.

The corresponding distribution for the SZE observable \ycyl\ is shown in the right panel of Fig. 2 for the same cluster. The A10 \ycyl-mass relation has been used to correct this distribution for redshift evolution.
 The same distributions are shown in black in the lower panels of Fig.~\ref{fig:multi_2}. 

In the following, we will refer to the joint probability of observing a particular value of $\zeta$ (or \ycyl) and \mvir\ evaluated from the SPT filtered data as $D_\textrm{SPT}(\zeta,M)$. For our purposes, $D_\textrm{SPT}(\zeta,M)$ represents therefore the \textit{true} SPT observable, from which we obtain \textit{derived} SZE observables by adopting external priors, as discussed in the next section.

We note that the effect of the SZE-optical central offset correction is larger for larger assumed cluster profiles. This  is a feature of the model adopted in S15, which describes the central-offset distribution normalized in terms of cluster virial radius \rvir. Therefore, for fixed cosmology and given a redshift, equal offsets in units of \rvir \, will be associated with larger angular offsets (and thus larger biases) for more massive clusters. 

\subsection{Breaking the Observable-Extent Degeneracy}

As a result of the degeneracy described in the previous section~and shown also in Fig.~\ref{fig:multi_1}, one needs extra information to extract the SZE observables. 
We first present the expectation for the SZE observable given richness \lam\ and then discuss how to estimate derived SZE observables 
for each RM cluster 
from the SPT filtered maps.

\subsubsection{Model Expectation}
\label{sec:model_exp}
\begin{figure*}
\centering { 
\hbox{
    \includegraphics[width=180mm]{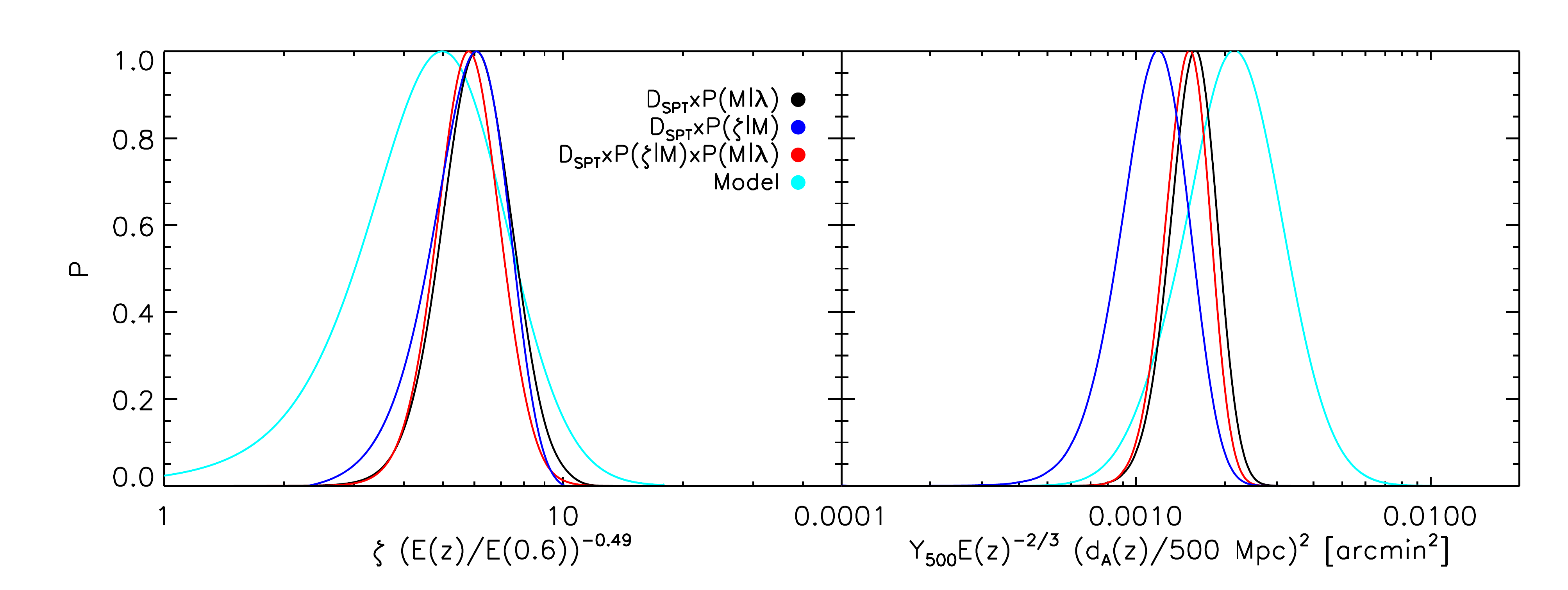}}
}
\caption{For the same example cluster shown in Fig.~\ref{fig:multi_1} and \ref{fig:multi_2}, we show the probability distribution function for the redshift-evolution-corrected  $\zeta$ (left) and \ycyls (right, for the A10 case) SZE observables and model predictions obtained by marginalizing the distributions shown in Fig.~\ref{fig:multi_2} over cluster extent \rvir. 
}
\label{fig:margin}
\end{figure*}

Given the two theoretical priors describing the richness-mass relation (red lines in Fig.~\ref{fig:multi_2}) and the SZE-mass relation (cyan lines in Fig.~\ref{fig:multi_2}) and under the assumption of uncorrelated intrinsic scatter, the expected distribution of $\zeta$ (and similarly for \ycyl) for a cluster selected with richness \lam\ at redshift $z$ can be expressed as:
\be
P^\textrm{Model}(\zeta|\lambda,z) = \int dM\,P(\zeta|M,z)P(M|\lambda,z).
\label{eq:model}
\ee
The resulting distributions are shown in the last panels (g and h) of Fig.~\ref{fig:multi_2} and as cyan lines after marginalizing over cluster extent in Fig.~\ref{fig:margin}.

We now discuss three approaches to break the degeneracy between observable and cluster extent, where simplified versions of the first two approaches have been previously adopted in the literature, and the third approach makes use of all information from both the optical and SZE observable mass scaling relations.

\subsubsection{Deriving Observables with \lam-mass Relation (Approach 1)}
\label{sec:approach1}
Following previous analyses \citep[e.g.][]{planck_opt,planck_lbg,hand11,drapper12,sehgal13}, the first approach is to break the degeneracy between the SZE observable and assumed cluster extent by fixing the angular scale of the cluster to the mean radius for a given richness with observed richness \lam\ and redshift $z$ using the richness implied mass $\langle M |\lambda,z \rangle$. Here, we improve and extend this approach by fully and explicitly including the prior on the scale of the cluster as:
\be
P^{App.1}(\zeta|\lam,z) = \int dM\, D_\textrm{SPT}(\zeta,M)P(M|\lambda,z).
\label{app1}
\ee
For the example cluster shown in Fig.~\ref{fig:multi_1}, this means we extract the measured observable by including the prior $P(M|\lambda,z)$ obtained following S15, represented by red lines in Fig.~\ref{fig:multi_2} as shown in panels a) and d). Note that, as mass maps directly into \rvir\, the associated distributions are therefore described by vertical lines 
in Fig.~\ref{fig:multi_2}.  The two resulting distributions for the two derived observables $P(\zeta|\lambda,z)$ and $P($\ycyl$|\lambda,z)$ are then obtained by marginalizing over the assumed cluster extent. These appear in the left and right panels of Fig.~\ref{fig:margin} for the $\zeta$ and \ycyls cases as solid black lines.

\subsubsection{Deriving Observables with SZE-mass Relation (Approach 2)}
\label{sec:approach2}
The second approach to break the degeneracy between SZE observable and cluster extent is to adopt an approach similar to the one described in \cite{gruen14} and \cite{planck15clusters}. Namely, we adopt the expected SZE observable-mass relations $\langle \zeta|M,z \rangle$ and $ \langle$\ycyl$|M,z \rangle$ and calculate the expected cluster extent for each value of $\zeta$ and $\ycyl$. In this approach we fully and explicitly include the priors $P(\zeta|$\rvir$,z)$ and $P($\ycyl$|$\rvir$,z)$ (respectively shown as cyan lines in the upper and bottom panels of Fig.~\ref{fig:multi_2} for the $\zeta$ and A10 \ycyl-mass scaling relation), obtained by the convolution of the assumed log-normal $\zeta$-mass (equation~\ref{eq:zeta}) and \ycyl-mass (equations~\ref{eq:arnaud10sc} and  \ref{eq:yx}) scaling relations with the Gaussian measurement uncertainties $\Delta \zeta ($\rvir$)$ and $\Delta$\ycyl(\rvir) as shown in panels b) and e).  The resulting distributions, marginalized over the cluster extent, are thus described by:
 \be P^{App.2}(\zeta|z) = \int dM\, D_\textrm{SPT}(\zeta,M)P(\zeta|M,z), \label{app2} \ee
and are shown in blue in the left and right panels of Fig.~\ref{fig:margin} for the $\zeta$ and A10 \ycyls cases, respectively.

\subsubsection{Deriving Observables with Combined Approach (1 and 2)}
\label{sec:approach3}
Finally, we note that given the two theoretical priors described above, the best estimator is obtained by combining all the available information, 
under the assumption that neither method is biased.
Specifically, given a model that describes both the richness-mass and SZE observable-mass relations 
with uncorrelated intrinsic scatter,
we adopt the combination of the above mentioned priors, as highlighted in panels c) and f) of Fig.~\ref{fig:multi_2}.
 
The resulting distributions, marginalized over the cluster extent, are thus described by:
\be P^\textrm{Comb}(\zeta|\lam,z) = \int dM\, D_\textrm{SPT}(\zeta,M)P(\zeta|M)P(M|\lambda,z), \label{app3} \ee  
and are shown in red in the left and right panels of Fig. \ref{fig:margin} for the $\zeta$ and for the A10 \ycyls cases, respectively. 

We note that the three SZE observable-extent constraints shown in panel c) of Fig.~\ref{fig:multi_2} cross in the same part of the $\zeta$-\rvir \, plane. As a result, the distributions for the three derived observables shown in the left panel of Fig.~\ref{fig:margin} nicely overlap. This is not surprising, as the \lam-mass relation and the $\zeta$-mass relation were both calibrated directly from the SPT-selected sample of clusters.  On the other hand, we note that this is not the case for the \ycyls observable, where the three SZE observable-extent constraints in panel f) of Fig.~\ref{fig:multi_2} overlap less.  As a consequence, marginalized distributions in the SZE observable for the three derived observables shown in the right panel of Fig.~\ref{fig:margin} are less consistent. We caution, however, that the derived observables are highly correlated with the model prediction as they share the same prior information.


\begin{figure}
\centering {\hbox{\includegraphics[width=90mm, height = 210mm,trim = 20 0 -20 0]{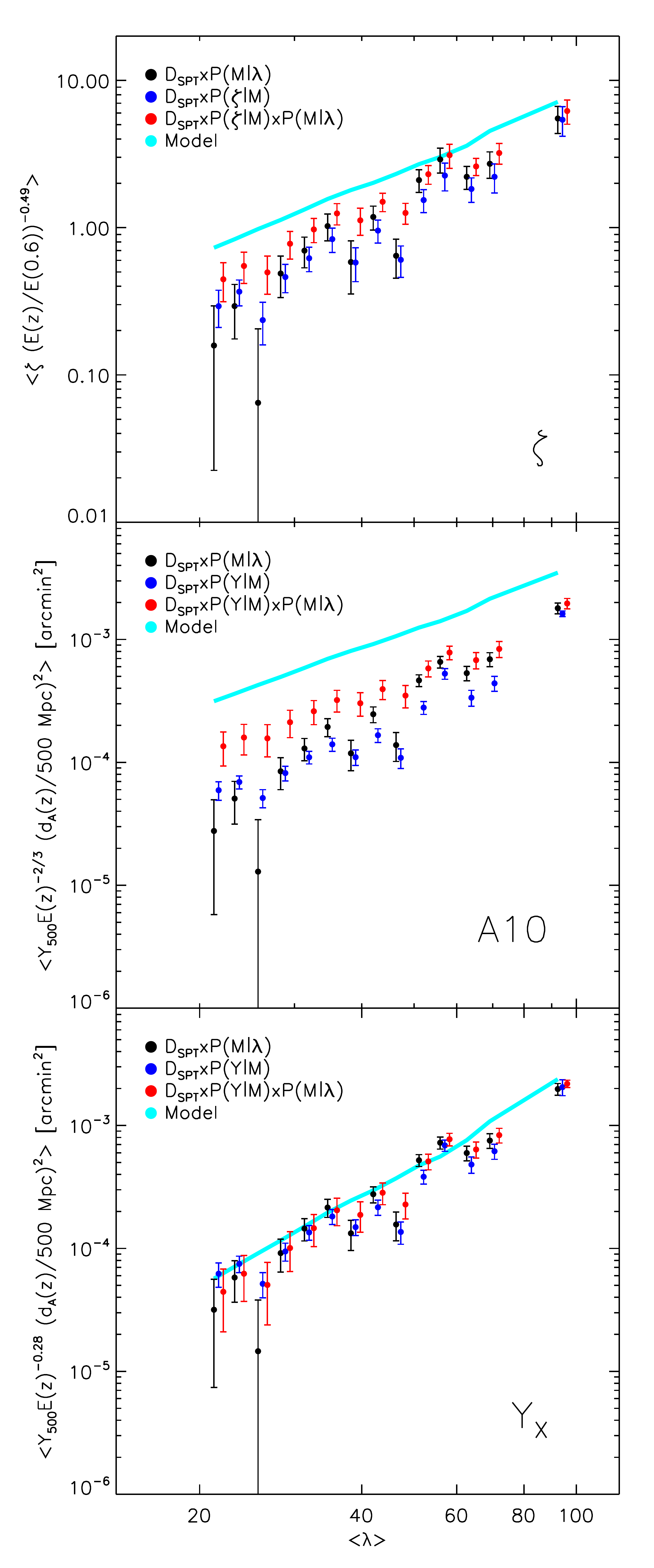}}}
\vskip-0.2in
\caption{Mean redshift-evolution-corrected SZE observables from clusters stacked in richness bins.  Results are for $\zeta$ (top), \ycyl\ assuming the A10 model (middle) and \ycyl\ derived from the best-fit SPT \yx-mass relation (bottom).  Data points show the average SPT-SZ SZE observables estimated including: a) a \lam-mass prior from S15 (black), b) an SZE observable-mass prior (blue), and c) both a and b (red).  The cyan line represents the mean expectation (equation \ref{eq:model}). A \lam\ dependent tension between the expectation and the observables exists to different degrees for all three scaling relations.  Formal evaluation of the tension in terms of either mass or observable biases is carried out using equation~(\ref{eq:lik}) and results appear in Fig.~\ref{fig:bias}.
\label{fig:y_l}}
\end{figure}

\begin{table*} 
\caption{Mean and $1\sigma$ uncertainty for the mass bias $b$ and observable contamination $f$ parameters as constrained through equation~(\ref{eq:lik}).}
\label{tab1}
\begin{tabular}{lcc|cc|ccccc}
 \lam\, Range & \multicolumn{2}{c}{$\zeta$-\lam} &\multicolumn{2}{c}{\ycyl-\lam\, A10} & \multicolumn{2}{c}{\ycyl-\lam\, SPT} \\ \hline\\[-7pt]
& $1-b$ & $1-f$& $1-b$ & $1-f$& $1-b$ & $1-f$ & \\ \hline\\[-7pt]
$\lambda >20$ & $0.74\pm0.07$&$0.59\pm0.12$&$0.52\pm0.05$&$0.31\pm0.08$&$0.84\pm0.07$&$0.71\pm0.14$&\\[3pt]
$\lambda >20$, no miscentering &$0.61\pm0.06$&$0.43\pm0.10$&$0.44\pm0.04$&$0.23\pm0.07$&$0.73\pm0.06$&$0.54\pm0.12$&\\[3pt]
$20<\lambda <40$ &$0.62\pm0.10$&$0.44\pm0.16$&$0.39\pm0.06$&$0.19\pm0.10$&$0.77\pm0.10$&$0.59\pm0.19$&\\[3pt]
$40<\lambda <80$ &$0.78\pm0.07$&$0.65\pm0.12$&$0.56\pm0.05$&$0.35\pm0.09$&$0.87\pm0.07$&$0.75\pm0.13$&\\[3pt]
$\lambda >80$ &$0.94\pm0.10$&$0.89\pm0.16$&$0.76\pm0.07$&$0.61\pm0.12$&$0.96\pm0.08$&$0.91\pm0.16$&\\[3pt]
\hline
\end{tabular}
\end{table*}

\begin{figure}
\centering { \hbox{\includegraphics[width=90mm,trim = 40 0 -40 0]{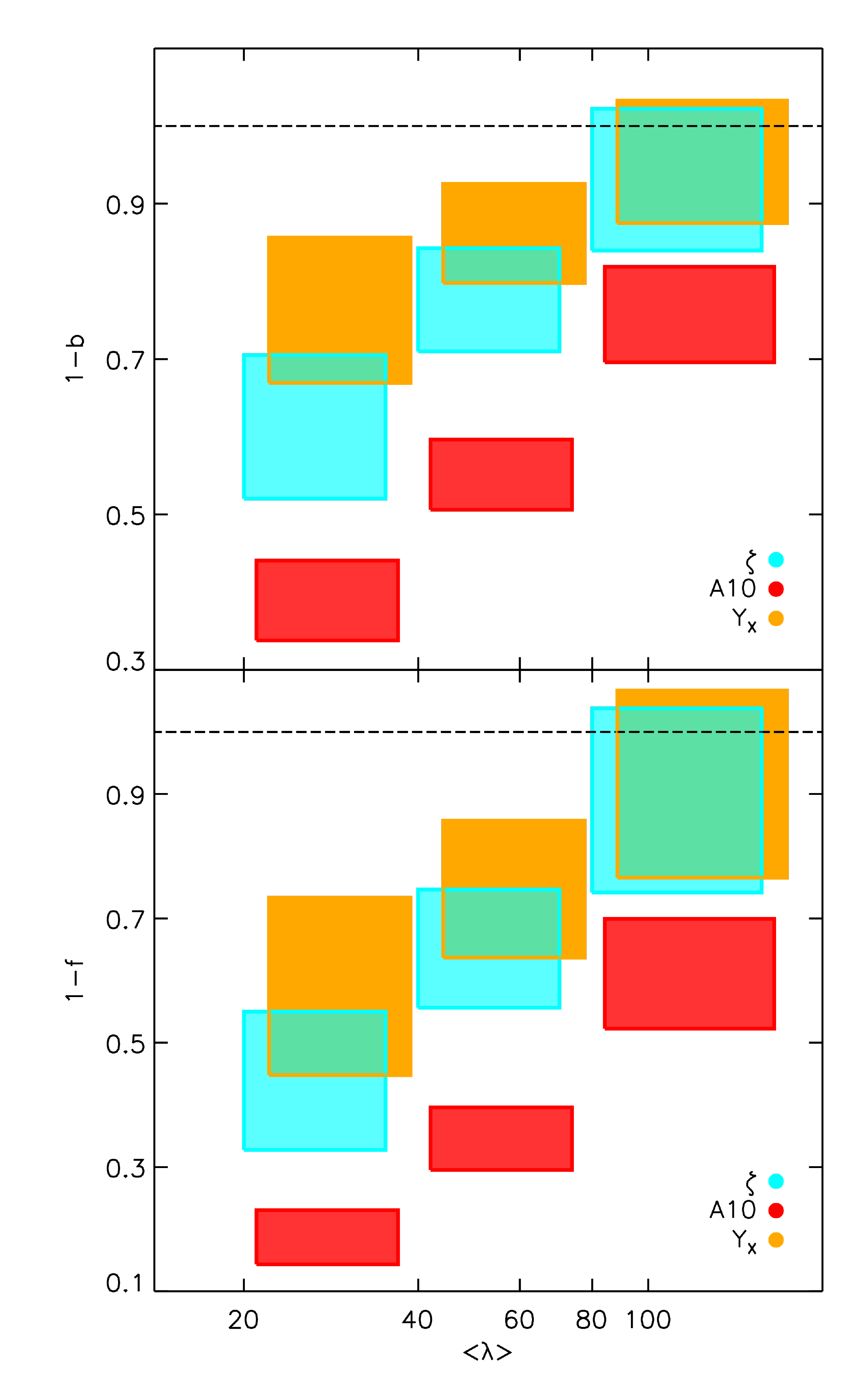}}
}
\caption{\textit{Upper panel}: Constraints on mass bias parameter $b$ obtained by evaluating equation~(\ref{eq:lik}) assuming no observable contamination ($f =0$) in three different richness bins. For each \lam-bin, $1\sigma$ regions of mass bias are shown for the $\zeta$-\lam\, relation (cyan), for the A10 derived \ycyl-\lam\, relation (red), and for the best-fit SPT derived \ycyl-\lam\, relation (yellow).  The regions have been slightly shifted in \lam\ for clarity. 
\textit{Bottom panel}: same as upper panel, but for the case of observable contamination parameter $f$ under the assumption of no mass bias ($b=0$). \label{fig:bias}}
\end{figure}

\section{Results}
\label{sec:results}
In this section we first present our measurements of three SZE observable-\lam\, relations and then discuss the consistency of these measurements with the model expectations. We comment on the impact of central-offset corrections in section \ref{sec:misc}.

\subsection{SZE Observable-Richness Relations}
\label{sec:model}

In the previous section, we demonstrated the extraction of $\zeta$ and \ycyl\ assuming three different priors that allow us to constrain the cluster radial extent. We now divide the RM sample into 14 equal spaced logarithmic bins of \lam\, and within each bin we compute the average SZE observables and model prediction.
The resulting average $\zeta$ corrected by the expected redshift evolution (with associated error bar) is shown in the upper panel of Fig.~\ref{fig:y_l} as a function of the mean \lam\ for the different methods of extracting $\zeta$.  These are color coded with Approach 1 (equation~\ref{app1}) in black, Approach 2 (equation~\ref{app2}) in blue and the Combined Approach (equation~\ref{app3}) in red.  The model expectation (equation~\ref{eq:model}) is the cyan line.   We find that the adopted model tends to overpredict the amplitude of $\zeta$ compared to the observed value, regardless of which of the three methods is used to extract $\zeta$. 

This result reflects our finding in S15, where the fraction of RM-selected clusters with SPT-SZ counterparts is found to be in mild tension with the expectation. This discrepancy could be due to one or both of contamination of the SZE observable through point sources \citep[i.e., radio galaxies or star forming galaxies; see, e.g.,][]{liu15} and contamination of the optical cluster sample. The contamination in the RM sample is expected to be at the $\sim 10 \pm 5\%$ level \citep{rykoff14}. In forthcoming papers, we will examine the contamination on the SZE side by studying deviations of the SZE spectrum with respect to the theoretical expectation for a sample of RM-selected clusters overlapping with the SPT-SZ survey (Bleem et al., in preparation) and by examining the 150~GHz radio galaxy luminosity functions in an ensemble of X-ray selected clusters \citep{gupta16}. In the next section we quantify the tension between the theoretical expectation and data in a rigorous statistical manner. 

The middle panel of Fig.~\ref{fig:y_l} shows the results obtained for the case of the A10 model for \ycyl. We find here that the adopted scaling relations (cyan) over-predicts the amount of SZE flux with respect to the all three SZE derived observables by a large factor, as we will discuss more in detail in section \ref{sec:5.2.2} . These results are qualitatively in agreement with previously published results using the maxBCG sample with \textit{Planck} \citep[e.g.,][]{planck_opt} and ACT \citep[e.g.,][]{sehgal13} data.  A quantitative comparison between our results and these works is complicated due to the different assumptions made for the richness-mass relation and due to the different sample properties. Focusing on the SZE estimator of Approach 1 (which is closer to the one adopted by \citealt{planck_opt} and \citealt{sehgal13})  represented by the black points, we also note that the slope of the derived \ycyl-\lam\ relation appears to be steeper than the model-predicted one, a result that is also found in the \textit{Planck} and ACT collaboration analyses. 

Finally in the bottom panel of Fig.~\ref{fig:y_l} we show the results obtained for the case of the \ycyl\ model derived from the best-fit \yx-mass relation self-consistently obtained from the SPT sample. Here, too, the signal is smaller than expected, but the difference is much smaller than for the A10-generated results and closer to the $\zeta$-\lam\, relation results shown in the top panel.  We quantitatively assess the consistency of measurements and expectations in the next section.

\subsection{Constraining Mass and Observable Bias Parameters}

To characterize a possible tension between the derived observables and the theoretical expectations, we extend the models and fit for the bias parameters $b$ and $f$ (equation \ref{eq:zeta}, \ref{eq:arnaud10sc}, \ref{eq:yx} ), where we are assuming either that the fractional mass bias $b$ is the same for all masses within a cluster subsample or that the fractional observable bias $f$ is the same for all observables within a cluster subsample. The probability $P$ for the 
SPT observable $D_\textrm{SPT}(\zeta,M)$ and bias parameters $b$ and $f$
is constructed by taking the product of the individual cluster probabilities $P_i$ as :
\be
P=\prod_{i=1}^{N_\textrm{clus}} P_i=\prod_{i=1}^{N_\textrm{clus}}\iint d\zeta dM\,D^i_\textrm{SPT}(\zeta,M) P(\zeta|M,f,b,z_i)P(M|\lambda_i,z_i).
\label{eq:lik}
\ee
We then constrain the mass bias parameter $b$ while assuming no observable contamination $f=0$ or we constrain the contamination parameter $f$ while assuming no mass bias $b=0$ for the tested models. We note also that equation~(\ref{eq:lik}) is similar to the Combined Approach above (equation~\ref{app3}), where $P(\zeta|M,f,b,z)$ is described by equation~(\ref{eq:zeta}). 

Introducing $b$ and $f$ allows us to explore the scale of mass biases or contamination that would be required to achieve consistency between the prediction and the derived observables. As we have defined them, values of these parameters consistent with zero are expected if there is no evidence of mass bias or no evidence of observable contamination.

\subsubsection{SPT-derived $\zeta$-$\lambda$ Relation}

For the whole sample used in this work ($\lam > 20$), and for the SZE observable $\zeta$ we find that the tested model with no bias or contamination is excluded at more than $3\sigma$. In particular, we find that the masses would have to be reduced by the factor $1-b=0.74 \pm 0.07$ or, equivalently, that the observable would have to be reduced by a factor $1-f = 0.59 \pm 0.12$ (Table~1).  Even then the fit would be poor, because as is clear in Fig.~\ref{fig:y_l} there is a \lam\, dependence in the offsets between the observables and the model.

In other words, the combination of the \lam-mass and $\zeta$-mass relations we calibrated in S15 using the high-mass SPT SZE-selected cluster sample is not a good description of the data. 
This is an indication that one or more of the underlying assumptions is invalid.  These assumptions include 1) fixing the cosmological parameters to particular values, 2) adopting the optical-SZE center offset distribution calibrated in S15, and 3) describing the SZE and optical mass--observable relations with a single power law with 
constant log-normal scatter over the full range in \lams that we explore.  Thus, perhaps this result is not entirely surprising. 

Given the strong cosmological constraints available today \citep{planck15cosm}, we do not expect marginalization over cosmology to have a significant impact; however, we will explore this explicitly in an upcoming analysis of a larger sample. The impact of corrections due to the assumed central-offset distribution model is discussed in section~\ref{sec:misc}. The analysis of S15 was limited to a sample of 19 clusters and did not allow for further exploration of the parameters space. Future works using data from the DES survey will greatly benefit from the larger statistics available and will allow us to better constrain and test the validity of the assumption adopted here.

To examine \lam\ dependent effects, we derive the best fit mass bias and observable contamination parameters $b$ and $f$ within three richness bins: $20<\lam<40$, $40<\lam<80$ and $\lam>80$. In this analysis, the error-bars on the three analyzed \lam\ bins are correlated through the marginalization of the \lam-mass scaling relation parameters. Results are reported in Table~1 and highlighted in Fig.~\ref{fig:bias} for the mass bias parameter $b$ and for the observable contamination parameter $f$ in the upper and lower panels, respectively. We note that when restricting this analysis to the highest richness bin $\lam >80$ (which is where the SPT SZE-selected clusters used to calibrate the observable mass relations lie) we obtain a bias parameter $1-b = 0.94\pm0.10$ ($1-f = 0.89 \pm 0.16$), consistent with no bias. Thus, in the \lam\, range where the \lam-mass and $\zeta$-mass were calibrated, there is no statistically significant evidence for bias or contamination, which is what we would expect. 

However, significant biases or contamination are associated with the lower richness bins, where the best fit parameters $b$ and $f$ are larger than zero with more than $3\sigma$ significance.
This suggests that a \lam-dependent bias would be needed to make the model prediction and observables consistent. For example, the lowest richness bin at $20<\lam<40$ can be described either with a mass bias of $1-b=0.62\pm0.10$ or with an observable contamination fraction of $1-f = 0.44\pm0.16$; most probably, the tension between the observables and model is due to a mix of different effects. As previously discussed, the origins of the tension could lie either in the RM catalogue 
through a richness dependent contamination or scatter, in the SZE estimator due to a mass dependent bias, e.g., star formation or radio loud AGN \citep[e.g.][]{lin09,sehgal10,liu15,lin15,planck15dust,gupta16} or perhaps in an SZE-optical center offset distribution that is much broader at low \lams than at high \lams where it was measured. The trend in Fig.~\ref{fig:bias} could also reflect increasing variance in halo mass selected at lower lambda which would produce lower mean expected SZE signal relative to the model used here \citep{evrard14}.  This effect could be amplified by an anti-correlation between galaxy and hot gas mass fractions, a feature seen in recent hydrodynamic simulations of the Rhapsody-G sample \citep{wu15}.  

A larger sample and stronger constraints from external mass calibration datasets and for the optical miscentering distribution will allow us to more easily unravel the remaining sources of tension.
We plan to pursue this in a future analysis.

\subsubsection{$\ycyl$-$\lambda$ Relation derived using the A10 Relation}
\label{sec:5.2.2}
For the SZE observable \ycyl\ associated with the A10 model, we derive $1-b=0.52 \pm 0.05$ ($1-f = 0.31 \pm 0.08$). 
This best fit mass bias parameter ($1 - b = 0.52$) is in good agreement with the best-fit mass hydrostatic mass bias parameter estimated by the \cite{planck15clus} ($1 - b = 0.58$) to reconcile the cosmological parameters fit to their CMB and cluster data sets.
We note, however, that a direct statistical comparison of the constraints on $1-b$ with the \textit{Planck} results is not straightforward for the following reasons. The adopted cosmology used to calibrate the scaling relations in S15 is similar but not identical to the \textit{Planck} CMB preferred cosmology (mostly due to differences in the amplitude of the power spectrum of density fluctuations). The mass and redshift range of the RM sample studied here is different from that of the \textit{Planck} cluster sample. A robust comparison between these results and the \textit{Planck} ones requires a detailed description of the contamination fraction in the RM sample studied here. Thus, the consistency between our constraints on the $1-b$ parameter and the ones reported by \cite{planck15clus} could well be coincidental.

We note that when only restricting this analysis to the sample of $\lam >80$ we obtain a bias parameter $1-b = 0.76 \pm 0.07$ ($1-f =  0.61 \pm 0.12$), consistent with theoretical expectations for hydrostatic mass bias \citep{nagai07,rasia12} and with recent weak lensing mass calibrations of \textit{Planck} clusters that indicate $ 1-b =  0.688 \pm 0.072$ \citep{vonderlinden14} and $1-b =  0.76 \pm 0.05$ (stat) $\pm 0.06$ (syst) \citep{hoekstra15},
but in mild tension with results reported from the LoCuSS sample: $1-b=0.95 \pm 0.04$ \citep{smith16}.
Also in this case, a strong richness-dependent bias is required to account for the different slope of the observed \ycyl-\lam\ relation and the expectation, as highlighted by Fig.~\ref{fig:bias}. Finally, note also that, contrary to the A10 \ycyl-mass relation, the SPT calibrated $\zeta$-mass and \yx-derived \ycyl-mass relations are not relying on the assumption of hydrostatic equilibrium, as they have been derived by matching the SPT clusters number counts with the reference cosmology.

\subsubsection{SPT-\yx-derived $\ycyl$-$\lambda$ Relation}

Finally, for the SPT derived \ycyl-\lams relation, we derive $1-b=0.84 \pm 0.07$ ($1-f = 0.71 \pm 0.14$).
In this case, the model prefers a mass-independent bias parameter that is closer to unity than that for the model describing the $\zeta$-mass relation. Our results exhibit a weak tension ($2.3\sigma$) with the model calibrated in S15.

When restricting this analysis to the sample with $\lam >80$, we obtain a bias parameter $1-b =0.96\pm0.08$ ($1-f = 0.91\pm 0.16$), consistent with no bias. While we find that $b=0$ is only weakly in tension ($\lesssim 3\sigma$) in the lowest \lam\ bin, a \lam-dependent bias is also preferred for this model (Fig.~\ref{fig:bias}).  As described previously, the parameters of the \lam-mass relation, the $\zeta$-mass relation, and the \ycyl-mass relation have been simultaneously calibrated from abundance matching at the reference cosmology using the SPT SZE-selected cluster sample. Therefore, we expect (and observe) consistency of these scaling relations in the richer RM clusters.

The smaller required bias in the SPT-derived \ycyl-\lams relation
in comparison with the bias required from the $\zeta$-\lams relation could indicate that the extrapolation of the \ycyl-mass relation calibrated at high masses to low mass clusters is providing a better description of the data than the same extrapolation with the $\zeta$-mass relation (see Fig.~\ref{fig:bias}).

\subsubsection{The Effects of Central-Offset Corrections}
\label{sec:misc}
Here we further discuss the impact of corrections due to the central-offset distribution in our measurements. As shown by \citet{sehgal13}, due to the shape of the optimal matched-filter, miscentering corrections have larger impacts in arcminute resolution experiments such as ACT and SPT as compared to the lower angular resolution \textit{Planck} experiment. Our results already include a correction for the effects of the SZE-optical central offset distribution that employs the model calibrated using a sample of 19 SZE-selected clusters and their RM counterparts (S15).  Results are summarized in Table~1. 

For the case of the $\zeta$-\lam\ relation, we note that if we had completely neglected the bias associated with the SZE-optical offset distribution $\beta$, we would increase the tension and obtain best fit bias parameters $1-b=0.61 \pm 0.06$ and $1-f = 0.43\pm 0.10$. For the A10 \ycyl-\mvir\ scaling relation, neglecting completely the bias associated with the SZE-optical offset distribution $\beta$ would require a bias parameter of $1-b = 0.44 \pm 0.04$ and $1-f = 0.23 \pm 0.07$. Finally, in the case of the SPT derived \ycyl-\mvir\ relation, neglecting completely the bias associated with the SZE-optical offset distribution would require bias parameters $1-b = 0.73\pm 0.06$ and $1-f = 0.54 \pm 0.12$.  

In essence, our adopted model for the SZE-optical central-offset distribution, which we calibrated on the high-\lam\ tail of the richness distribution represented by the SPT-selected sample, is shifting our constraints on the bias parameters closer to the expectations by approximately $2\sigma$.  As discussed, however, there is a preference in each of these scaling relations for a richness-dependent bias or contamination;  an interesting possibility is that the typical SZE-optical central-offsets could be a significantly larger fraction of the cluster extent at low \lams than at high \lam, where we have measured them.  A sample of SZE selected clusters extending to much lower mass would allow the SZE-optical central-offset distribution to be directly constrained at low \lam.

\section{Conclusions}
\label{sec:conclusions}
In this study we examine galaxy cluster SZE-optical scaling relations and the underlying SZE observable-mass and optical richness \lam-mass relations.  We extend our study beyond the high mass, SPT SZE-selected sample studied in S15 to include systems of lower mass and richness where the SZE observables of individual clusters have much lower amplitude.  To do this we use a sample of 719 optically selected RM clusters \citep{rykoff16} with $\lam>20$ that have been selected from a 124.6~\degs\ region of the DES-SV dataset with overlapping SPT-SZ mm-wave data.   At the locations of the optically selected clusters we extract the SZE observables from the SPT-SZ maps and stack these signals within richness bins to constrain the mean SZE observable as a function of \lam.  The SZE observables we measure are SPT detection significance $\zeta$ and integrated Compton-$y$ \ycyl, allowing us to study the $\zeta$-\lams relation and two different \ycyl-\lams relations.

We show in our matched-filter analysis of the mm-wave maps that the derived SZE observable associated with each RM-selected cluster depends on the assumed cluster extent or virial radius.  
To determine the range of cluster virial radius that is relevant for each cluster, we adopt priors for the SZE observable-mass relation and/or for the \lam-mass relation. 
We extract observables in each cluster using three different approaches: (1) a prior on the richness-mass relation $P(M|\lambda,z)$ which specifies the cluster extent \rvir\ (equation~\ref{app1}); (2) a prior on the SZE observable-mass relations $P(\zeta|M,z)$ or $P($\ycyl$|M,z)$ (equation~\ref{app2}); and (3) the combination of the two $P(\zeta|M,z) \times P(M|\lambda,z)$ and $P($\ycyl$|M,z) \times  P(M|\lambda,z)$ (equation~\ref{app3}).  In all cases we marginalize over the full probability distribution function associated with the adopted priors.  This leads to three somewhat different measurements for each SZE observable in each RM-selected cluster.  Also, because any mismatch in the optical and SZE centers of the clusters will impact the extracted SZE observables, we correct for this effect using the optical-SZE central offset distribution measured in S15.

We then stack the SZE observables of the RM-selected sample in 14 bins of richness  \lam, calculating the average amplitude and uncertainty of the SZE observable in each bin.  Our stacking technique explicitly propagates the prior information into the derived SZE observable amplitude and uncertainty. We use these data to construct the three SZE-optical scaling relations:  1) the $\zeta$-\lams relation, 2) the \ycyl-\lams relation adopting the A10 prior and 3) the \ycyl-\lams relation from SPT.  For each relation we then examine the consistency of the observed relation with the expectation, finding that there is poor agreement between prediction and data for all three relations (see Fig.~\ref{fig:y_l}).  In all cases the agreement is better at high \lams than at low.  

We explore the scale of the tension between expectation and observations by adopting either a mass bias parameter $b$ or an observable contamination parameter $f$. This contamination could be associated with a higher degree of SZE contamination due to unresolved star formation and radio galaxies associated with low-richness clusters or to a higher degree of contamination in the RM sample at lower \lams \citep{liu15,gupta16}. Both bias parameters are integrated into the SZE observable-mass relations (see section~\ref{sec:theory}) and can be extracted for the full RM sample or subsets of the sample.  Results are summarized below.

For the $\zeta$-\lams relation (equation~\ref{eq:zeta}, upper panel in Fig.~\ref{fig:y_l}) we find that the model calibrated through the SPT selected sample in S15 tends to over-predict the signal of the derived observables. This tension can be alleviated by introducing a mass bias factor of $1-b=0.74 \pm 0.06$, which could be also explained through a contamination factor $1-f = 0.59$, which could model the combination of SZE observable contamination or contamination of the RM catalog due to projection effects.  However, the slope of the expected $\zeta$-\lams relation is significantly shallower than the observed one, and the bias or contamination would have to be more pronounced for low richness systems than for high richness systems (see Fig.~\ref{fig:bias}). An analysis of the richest $\lam>80$ clusters shows good consistency between the data and the expectation, supporting a picture where contamination effects of either the SZE observable or the RM catalog (or a combination of these effects) are larger for lower mass systems. Alternatively, the adopted model for the SZE-optical center offset corrections may not properly describe the data for the lower richness clusters
or the intrinsic scatter in the $\lam$-mass relation may be larger at low $\lam$ \citep[][]{evrard14}.

For the \ycyls-\lams relation that adopts the A10 based \ycyl-mass scaling relation calibration (equation~\ref{eq:arnaud10sc}, middle panel of Fig.~\ref{fig:y_l}), the expectation prediction is higher than the observations by a factor of $\sim3$. This inconsistency can be reduced by adopting a mass bias parameter $1-b = 0.52 \pm 0.05$ or an observable contamination factor of $1-f = 0.31 \pm 0.08$.  This best fit mass bias parameter $1-b = 0.52$ is in good agreement with the best-fit mass bias parameter $1-b=0.58$ required to reconcile the cosmological parameters obtained through analysis of the \textit{Planck} CMB anisotropy with the cosmological parameters derived from the \textit{Planck} cluster sample when adopting an XMM hydrostatic mass calibration. We caution, however, that this result could be largely coincidental, because the slope of the best fit relation for the measured observables is significantly steeper than the slope of the expectation, and measurements again indicate the need for a richness dependent contamination or mass bias (see Fig.~\ref{fig:bias}). The analysis of the richest $\lam>80$ subsample results in a best fit mass bias parameter $1-b=0.76\pm 0.07$, which is smaller and is consistent with estimates of the hydrostatic mass bias from simulations and from weak lensing measurements \citep{nagai07,rasia12,vonderlinden14,hoekstra15}, 
but is in mild tension with the weak lensing measurements from LoCuSS \citep{smith16}.
Contamination effects that are \lams dependent would be similar to those described above.

For the \ycyl-mass relation derived 
from the SPT sample (equation~\ref{eq:yx}, bottom panel of Fig.~\ref{fig:y_l}), we derive a mass bias factor $1-b=0.84 \pm 0.07$ and an observable contamination factor $1-f = 0.75\pm 0.13$, which are smaller biases than those in the $\zeta$-\lams relation and marginally in tension with zero mass bias. The larger bias required for the $\zeta$-\lams relation could be caused by a break in the relation for lower mass systems, which perhaps is not required in the \ycyl-\lams relation.  When restricting this analysis to the sample at $\lam >80$ we obtain a bias (contamination) parameter $1-b =0.96 \pm 0.08 $ ($1-f = 0.91 \pm 0.16$), which is consistent with no bias. Also in this relation the slope is steeper in the measurements than in the expectation,  indicating that a \lam-dependent bias or contamination at the $\lesssim 3\sigma$ level is preferred (see Fig.~\ref{fig:bias}). 

Future work benefiting from the larger region of overlap between the DES and SPT surveys will further test the consistency between the scaling relations derived from the SPT selected sample and from the RM-selected clusters, and to test the validity of the models adopted in this analysis. Ultimately, a simultaneous calibration of cosmological parameters and of the richness-mass and SZE observable-mass scaling relations from the RM-selected sample will allow us to more precisely estimate tensions between the two samples and help in providing insights into the underlying causes.

\section*{Acknowledgements}

We acknowledge the support of the DFG Cluster of Excellence ``Origin and Structure of the Universe'', the Transregio program TR33 ``The Dark Universe'' and the Ludwig-Maximilians-Universit\"at. 
This paper has gone through internal review by the DES and SPT collaborations.  

The South Pole Telescope is supported by the National Science Foundation through grant PLR-1248097. Partial support is also provided by the NSF Physics Frontier Center grant PHY-1125897 to the Kavli Institute of Cosmological Physics at the University of Chicago, the Kavli Foundation and the Gordon and Betty Moore Foundation grant GBMF 947.

We are grateful for the extraordinary contributions of our CTIO colleagues and the DECam Construction, Commissioning and Science Verification
teams in achieving the excellent instrument and telescope conditions that have made this work possible. The success of this project also 
relies critically on the expertise and dedication of the DES Data Management group.
Funding for the DES Projects has been provided by the U.S. Department of Energy, the U.S. National Science Foundation, the Ministry of Science and Education of Spain, 
the Science and Technology Facilities Council of the United Kingdom, the Higher Education Funding Council for England, the National Center for Supercomputing 
Applications at the University of Illinois at Urbana-Champaign, the Kavli Institute of Cosmological Physics at the University of Chicago, 
the Center for Cosmology and Astro-Particle Physics at the Ohio State University,
the Mitchell Institute for Fundamental Physics and Astronomy at Texas A\&M University, Financiadora de Estudos e Projetos, 
Funda{\c c}{\~a}o Carlos Chagas Filho de Amparo {\`a} Pesquisa do Estado do Rio de Janeiro, Conselho Nacional de Desenvolvimento Cient{\'i}fico e Tecnol{\'o}gico and 
the Minist{\'e}rio da Ci{\^e}ncia, Tecnologia e Inova{\c c}{\~a}o, the Deutsche Forschungsgemeinschaft and the Collaborating Institutions in the Dark Energy Survey. 
The Collaborating Institutions are Argonne National Laboratory, the University of California at Santa Cruz, the University of Cambridge, Centro de Investigaciones Energ{\'e}ticas, 
Medioambientales y Tecnol{\'o}gicas-Madrid, the University of Chicago, University College London, the DES-Brazil Consortium, the University of Edinburgh, 
the Eidgen{\"o}ssische Technische Hochschule (ETH) Z{\"u}rich, 
Fermi National Accelerator Laboratory, the University of Illinois at Urbana-Champaign, the Institut de Ci{\`e}ncies de l'Espai (IEEC/CSIC), 
the Institut de F{\'i}sica d'Altes Energies, Lawrence Berkeley National Laboratory, the Ludwig-Maximilians Universit{\"a}t M{\"u}nchen and the associated Excellence Cluster Universe, 
the University of Michigan, the National Optical Astronomy Observatory, the University of Nottingham, The Ohio State University, the University of Pennsylvania, the University of Portsmouth, 
SLAC National Accelerator Laboratory, Stanford University, the University of Sussex, Texas A\&M University, and the OzDES Membership Consortium.
The DES data management system is supported by the National Science Foundation under Grant Number AST-1138766.
The DES participants from Spanish institutions are partially supported by MINECO under grants AYA2012-39559, ESP2013-48274, FPA2013-47986, and Centro de Excelencia Severo Ochoa SEV-2012-0234.
Research leading to these results has received funding from the European Research Council under the European Union’s Seventh Framework Programme (FP7/2007-2013) including ERC grant agreements 
 240672, 291329, and 306478.

\bibliographystyle{mnras}
\bibliography{paper,spt}
\textit{$^{1}$Faculty of Physics, Ludwig-Maximilians-Universit\"at, Scheinerstr. 1, 81679 Muenchen, Germany\\
$^{2}$Excellence Cluster Universe, Boltzmannstr.\ 2, 85748 Garching, Germany\\
$^{3}$Max Planck Institute for Extraterrestrial Physics, Giessenbachstrasse, 85748 Garching, Germany\\
$^{4}$Department of Physics, University of Arizona, 1118 E 4th St, Tucson, AZ 85721\\
$^{5}$Fermi National Accelerator Laboratory, P. O. Box 500, Batavia, IL 60510, USA\\
$^{6}$Kavli Institute for Cosmological Physics, University of Chicago, Chicago, IL 60637, USA\\
$^{7}$Department of Astronomy and Astrophysics,University of Chicago, 5640 South Ellis Avenue, Chicago, IL 60637\\
$^{8}$Kavli Institute for Particle Astrophysics \& Cosmology, P. O. Box 2450, Stanford University, Stanford, CA 94305, USA\\
$^{9}$SLAC National Accelerator Laboratory, Menlo Park, CA 94025, USA\\
$^{10}$Argonne National Laboratory, 9700 South Cass Avenue, Lemont, IL 60439, USA\\
$^{11}$Cerro Tololo Inter-American Observatory, National Optical Astronomy Observatory, Casilla 603, La Serena, Chile\\
$^{12}$Department of Physics \& Astronomy, University College London, Gower Street, London, WC1E 6BT, UK\\
$^{13}$Department of Physics and Electronics, Rhodes University, PO Box 94, Grahamstown, 6140, South Africa\\
$^{14}$Department of Physics, University of Chicago, Chicago, IL, USA 60637\\
$^{15}$Kavli Institute for Particle Astrophysics and Cosmology, Stanford University, 452 Lomita Mall, Stanford, CA 94305\\
$^{16}$Department of Physics, Stanford University, 382 Via Pueblo Mall, Stanford, CA 94305\\
$^{17}$CNRS, UMR 7095, Institut d'Astrophysique de Paris, F-75014, Paris, France\\
$^{18}$Sorbonne Universit\'es, UPMC Univ Paris 06, UMR 7095, Institut d'Astrophysique de Paris, F-75014, Paris, France\\
$^{19}$Laborat\'orio Interinstitucional de e-Astronomia - LIneA, Rua Gal. Jos\'e Cristino 77, Rio de Janeiro, RJ - 20921-400, Brazil\\
$^{20}$Observat\'orio Nacional, Rua Gal. Jos\'e Cristino 77, Rio de Janeiro, RJ - 20921-400, Brazil\\
$^{21}$Department of Astronomy, University of Illinois, 1002 W. Green Street, Urbana, IL 61801, USA\\
$^{22}$National Center for Supercomputing Applications, 1205 West Clark St., Urbana, IL 61801, USA\\
$^{23}$Institut de Ci\`encies de l'Espai, IEEC-CSIC, Campus UAB, Carrer de Can Magrans, s/n,  08193 Bellaterra, Barcelona, Spain\\
$^{24}$Institut de F\'{\i}sica d'Altes Energies (IFAE), The Barcelona Institute of Science and Technology, Campus UAB, 08193 Bellaterra (Barcelona) Spain\\
$^{25}$Institute of Cosmology \& Gravitation, University of Portsmouth, Portsmouth, PO1 3FX, UK\\
$^{26}$School of Physics and Astronomy, University of Southampton,  Southampton, SO17 1BJ, UK\\
$^{27}$Department of Astronomy, University of Michigan, Ann Arbor, MI 48109, USA\\
$^{28}$Department of Physics, University of Michigan, Ann Arbor, MI 48109, USA\\
$^{29}$Institute of Astronomy, University of Cambridge, Madingley Road, Cambridge CB3 0HA, UK\\
$^{30}$Department of Physics, University of California, Berkeley, CA 94720\\
$^{31}$Australian Astronomical Observatory, North Ryde, NSW 2113, Australia\\
$^{32}$Departamento de F\'{\i}sica Matem\'atica,  Instituto de F\'{\i}sica, Universidade de S\~ao Paulo,  CP 66318, CEP 05314-970, S\~ao Paulo, SP,  Brazil\\
$^{33}$George P. and Cynthia Woods Mitchell Institute for Fundamental Physics and Astronomy, and Department of Physics and Astronomy, Texas A\&M University, College Station, TX 77843,  USA\\
$^{34}$Kavli Institute for Astrophysics and Space Research, Massachusetts Institute of Technology, 77 Massachusetts Avenue, Cambridge, MA 02139\\
$^{35}$Department of Astrophysical Sciences, Princeton University, Peyton Hall, Princeton, NJ 08544, USA\\
$^{36}$Instituci\'o Catalana de Recerca i Estudis Avan\c{c}ats, E-08010 Barcelona, Spain\\
$^{37}$Jet Propulsion Laboratory, California Institute of Technology, 4800 Oak Grove Dr., Pasadena, CA 91109, USA\\
$^{38}$School of Physics, University of Melbourne, Parkville, VIC 3010, Australia\\
$^{39}$Department of Physics and Astronomy, Pevensey Building, University of Sussex, Brighton, BN1 9QH, UK\\
$^{40}$Centro de Investigaciones Energ\'eticas, Medioambientales y Tecnol\'ogicas (CIEMAT), Madrid, Spain\\
$^{41}$Department of Physics and Astronomy, University of Pennsylvania, Philadelphia, PA 19104, USA\\}

\end{document}